\documentclass[12pt,a4paper]{article}
\usepackage{amssymb,amsmath,latexsym}
\title{Boundary maps for $C^*$-crossed products with $\RR$ \\
with an application to the quantum Hall effect} 

\author{J. Kellendonk$^{1}$,
H. Schulz-Baldes$^{2}$,
\\
\\
\\
$^1$ {\small School of Mathematics, Cardiff University, 
Cardiff, CF24 4YH, Wales}
\\
$^2$ {\small Institut f\"ur Mathematik, TU Berlin, 
Strasse des 17. Juni 136, 10623 Berlin, Germany}
}

\date{\today}

\newtheorem{thm}{Theorem}
\newtheorem{defn}{Definition}
\newtheorem{prop}{Proposition}
\newtheorem{lem}{Lemma}

\newtheorem{ex}{Example}

\newcommand{\Real}{\mathbb R}
\newcommand{\RR}{\Real}
\newcommand{\Rh}{\hat{\Real}}
\newcommand{\Nat}{\mathbb N}
\newcommand{\Complex}{\mathbb C}
\newcommand{\HH}{\mathbb H_3}
\newcommand{\CC}{\Complex}
\newcommand{\Z}{\mathbb Z}

\newcommand{\Ea}{{\mathcal E}}
\newcommand{\Ta}{{\mathcal A}}
\newcommand{\Aa}{{\Ta_\infty}}

\newcommand{\aco}{\idl}
\newcommand{\aca}{\beta^{\|}}
\newcommand{\acb}{\beta^\perp}

\newcommand{\Uu}{{\cal U}}

\renewcommand{\H}{{\mathcal H}}

\newcommand{\PP}{{\bf P}}

\newcommand{\Bb}{{\mathcal B}}

\newcommand{\Tr}{\mbox{\rm Tr}}  
\newcommand{\TV}{{\mathcal T}}
\newcommand{\TVh}{\hat{{\mathcal T}}}
\newcommand{\tr}{\mbox{tr}}

\newcommand{\Cc}{{\mathcal C}}
\newcommand{\Jj}{{\mathcal J}}

\newcommand{\id}{{\mbox{\rm id}}}
\newcommand{\idl}{{\mbox{\rm\tiny id}}}
\newcommand{\eva}{{\mbox{\rm ev}}}
\newcommand{\eval}{{\mbox{\rm\tiny ev}}}

\newcommand{\bew}{{\bf Proof:}}
\newcommand{\eb}{\hfill $\Box$}

\newcommand{\x}{{\vec x}}
\newcommand{\y}{{\vec y}}

\newcommand{\hull}{\Omega}
\newcommand{\om}{\omega}
\newcommand{\oh}{{\hat{\om}}}

\renewcommand{\H}{{\mathcal H}}

\newcommand{\erz}[1]{\langle{#1}\rangle}
\newcommand{\pair}[2]{\erz{#1,#2}}
\newcommand{\diag}[2]{\mbox{\rm diag}(#1,#2)}
\newcommand{\tOmega}{{\Omega^s}}
\newcommand{\td}{{d^s}}
\newcommand{\tint}{{\int^s}}
\newcommand{\ttau}{{\tilde\tau}}

\newcommand{\hotimes}{{\hat\otimes}}
\newcommand{\K}{{\mathcal K}}

\newcommand{\CA}{$C^*$-algebra}
\newcommand{\CF}{$C^*$-field}

\newcommand{\im}{\mbox{\rm im\,}}

\newcommand{\cp}{{\rtimes}}

\renewcommand{\Cc}{{\mathcal C}}

\newcommand{\dom}{\mbox{\rm dom}}
\newcommand{\fin}{\mbox{\rm\tiny fin}}

\newcommand{\enn}{\mu}
\newcommand{\ch}{\mbox{\rm ch}}
\newcommand{\supp}{\mbox{\rm supp}}
\newcommand{\chd}{\ch}

\begin{document}

\maketitle

\begin{abstract}
The boundary map in $K$-theory arising from the
Wiener-Hopf extension of a crossed product algebra with $\RR$
is the Connes-Thom isomorphism. In this article the Wiener Hopf extension is
combined with the Heisenberg group algebra to provide
an elementary construction of a corresponding map on higher traces
(and cyclic cohomology). 
It then follows directly from a non-commutative Stokes
theorem that this map is dual w.r.t.\
Connes' pairing of cyclic cohomology with $K$-theory. 
As an application, we prove equality of quantized bulk and
edge conductivities for the integer quantum Hall effect described by 
continuous magnetic Schr\"odinger operators.  
\end{abstract}


\section{Motivation and main result}

In a commonly used approach to study aperiodic solids,
particles in the bulk of the medium are described by covariant
families of one-particle 
Schr\"odinger operators $\{H_\om\}_{\om\in\hull}$ where
$\hull$ is the probability space of configurations furnished with an
ergodic action of space translations. 
Crossed product algebras provide a natural framework for such
families \cite{Bel85}. 
In particular their bounded functions are represented by elements
of a $C^*$-crossed product, the so-called bulk algebra. 
The non-commutative topology of the \CA\
is a useful tool to construct topological invariants resulting from
pairings  between $K$-group elements and higher traces. 
Some of these invariants may be physically interpreted as topologically
quantised quantities; the quantised Hall conductivity is such an
example. The physics near a boundary of the solid can 
also be described by a \CA, the so-called edge algebra. The bulk
algebra being essentially a crossed product of the edge algebra with
$\RR$ (or with $\Z$ in the tight binding approximation \cite{KRS02}),
both algebras are tied together in the Wiener-Hopf extension (or
respectively the Toeplitz extension). This extension gives rise to
boundary maps between the $K$-groups and the higher traces of the bulk
and edge algebra which allow one to equate bulk and edge invariants. 
This topological relation 
and its physical interpetations is our main objective. 
We discuss one prominent
physical example of this, the quantum Hall effect, where the
Hall conductivity may either be expressed as the Chern number of a
spectral projection associated with a gap in the bulk spectrum 
\cite{Bel85,AS85,K87,ASS,Bellissard88,BES}
or as the non-commutative winding number of the unitary of time translation of
the edge states corresponding to the gap
by a characteristic time (the inverse of the gap width) \cite{KRS02,KS1}.   
The mathematical background for this equality between bulk 
and edge invariants for continuous Schr\"odinger
operators is the subject of the present article.

\vspace{.2cm}

The mathematical framework is as follows.
Consider an $\RR$-action $\alpha$ on a \CA\ $\Bb$.
Denote by $\tau$ the translation action of $\RR$ on the half open space
$\RR\cup\{+\infty\}$ (with fixed point $+\infty$). 
This defines a crossed product \CA\ $\Bb\rtimes_\alpha\RR$ 
and an extension of this \CA\
by another crossed product, $C_0(\RR\cup\{+\infty\},\Bb)
\rtimes_{\tau\otimes\alpha}\RR$, the so-called Wiener-Hopf
extension. They form an exact sequence 
\begin{equation}
\label{eq-introext}
0
\;\longrightarrow\;
\K \otimes \Bb
\;\longrightarrow\;
C_0(\RR\cup\{+\infty\},\Bb)
\rtimes_{\tau\otimes\alpha}\RR
\;\stackrel{\eva_{\infty}}{\longrightarrow}\;
\Bb\rtimes_\alpha\RR
\;\longrightarrow\;
0
\;
\end{equation}
where $\eva_{\infty}$ is induced from the surjective homomorphism
$C_0(\RR\cup\{+\infty\},\Bb)\to \Bb$ given by 
evaluating $f\in C_0(\RR\cup\{+\infty\},\Bb)$ at $+\infty$
and $\K$ are the compact operators on $L^2(\RR)$. 
Rieffel has shown \cite{Rieffel82} that the boundary maps
$\partial_i:K_i(\Bb\rtimes_\alpha\RR)\to K_{i+1}(\Bb)$
in the corresponding six-term exact sequence are the inverses of the
Connes-Thom isomorphism \cite{Connes81}. In the physical context
described above, the boundary maps relate the $K$-groups of the bulk
algebra with the $K$-groups of the edge algebra.

\vspace{.2cm}

In the context of smooth crossed products, where $\Bb$ is a Fr\'echet
algebra with smooth action $\alpha$ so that one obtains a smooth
version of (\ref{eq-introext}), Elliott, Natsume and Nest \cite{ENN1} have 
given dual boundary maps for cyclic cohomology groups, namely isomorphisms
$\#_\alpha:HC^n(\Bb)\to HC^{n+1}(\Bb\rtimes_\alpha\RR)$ which satisfy
\begin{equation}\label{eq-duality}
\pair{\#_\alpha\eta}{x}
\;=\;
-\,\frac{1}{2\pi}\;\pair{\eta}{\partial_i x}
\;,
\qquad \eta\in HC^{i-1+2n}(\Bb)\;,
\qquad x\in K_i(\Bb\rtimes_\alpha\RR)
\;,
\end{equation}
where $\pair{\cdot}{\cdot}$ denotes Connes' pairing between cyclic
cocycles and $K$-group elements.

\vspace{.2cm}

Our aim here is to obtain the same kind of result for
$\alpha$-invariant higher traces on 
\CA s. 
One reason for doing this is that, whereas
our estimates from \cite{KS1} show
that the operators relevant in the physical context described above
lie in $C^*$-crossed products
it is not clear whether they belong to the smooth sub-algebras
used in \cite{ENN1}. 
Another reason is to present a different proof with, as we believe, 
considerably simpler algebraic constructions so that it should
henceforth be easier accessible also to the non-expert. In fact,
for the phsyical interpretation of equation 
(\ref{eq-duality}) it is indispensible that all isomorphisms involved
can be made explicit. In particular, it is essential that we can
compute the boundary map $\partial_0$ on (classes of) spectral projections of
the Schr\"odinger operator on gaps. Our proof establishes (\ref{eq-duality})
directly for $i=0$, the case needed for the application to the quantum
Hall effect,
whereas the equality is proven in \cite{ENN1}
first for $i=1$ and then extended to $i=0$ using Takai
duality and Connes' Thom isomorphism. Hence the non-expert reader can
understand our result without prior knowledge, for instance, of Connes'
Thom isomorphism.
The proof we present uses continuous fields 
of \CA s and is inspired by another 
article of Elliott, Natsume and Nest \cite{ENN2}. 

\vspace{.2cm}

More precisely, the result can be described as follows. An $n$-trace
on a Banach algebra $\Bb$ 
is the character of an (unbounded) $n$-cycle $(\Omega,\int,d)$ over 
$\Bb$ having further continuity properties  ({\sl cf.} Def.~\ref{def3}). 
It is called $\alpha$-invariant if $\alpha$ extends to an action of $\RR$
on the graded differential algebra $(\Omega,d)$ by isomorphisms of
degree $0$ and $\int\!\circ\alpha=\int$ (and the abovementioned
continuity properties are $\alpha$-invariant, {\sl cf.} Def.~\ref{def4}).
Let $\eta$ be an $n$-trace which is the character of 
an $\alpha$-invariant cycle $(\Omega,\int,d)$ over $\Bb$. 
We prove that
$$
\#_\alpha\eta(f_0,\ldots,f_{n+1})
\;=\;
\sum_{k=1}^{n+1}\;(-1)^k\;\int 
\left(f_0d f_1\cdots\nabla f_k\cdots df_{n+1}\right)(0)
\;, \qquad \nabla f(x)=\imath x f(x)\;,
$$
is an $n+1$-trace on the $L^1$-crossed product $L^1(\RR,\Bb,\alpha)$.
Furthermore, if $\Bb$ is a \CA\ then the pairing with  $\#_\alpha\eta$
extends to the $K$-group of the $C^*$-crossed product $\Bb\rtimes_\alpha\RR$
and satisfies the duality equation (\ref{eq-duality}).

\vspace{.2cm}

Sections~\ref{sec-C} to \ref{sec-pairing} are devoted to explain the
mathematical context and to prove the above result 
(Theorem~\ref{prop2} and Theorem~\ref{theo-duality}). 
Theorem~\ref{theo-duality} follows from two main arguements, a homotopy
arguement (Theorem~\ref{gleich}) and periodicity in cyclic cohomology.
Although the latter is well-known we have added a detailed proof of its
version adapted to our context (Theorem~\ref{gleich2}) in the
appendix, hence making this work self-contained. 
In Section~\ref{sec-equal} we discuss the application of this result 
to the quantum Hall effect.

\section{$C^*$-algebraic preliminaries}
\label{sec-C}

\subsection{Crossed products by $\RR$}\label{appendix1}

Let $\alpha:\RR\to \mbox{\rm Aut}(\Bb)$ be an action of $\RR$ on a \CA\ $\Bb$. 
It is required to be continuous in the
sense that for all $A\in\Bb$, the function $x\in\RR\mapsto \alpha_x(A)$ is
continuous. 
The crossed product algebra $\Bb\rtimes_\alpha\RR$ of $\Bb$ 
with respect to the action $\alpha$ of $\RR$ is defined as follows \cite{Ped}.
The linear space $C_c(\RR,\Bb)$ of compactly supported
continuous functions with values in $\Bb$ is endowed with
the $*$-algebra structure
\begin{equation}
\label{prod}
(fg)(x)\; =\; \int_{\RR} dy \,f(y)\,\alpha_y(g(x-y))
\;,\qquad
f^*(x) \;=\; {\alpha_x (f(-x))}^*\;.
\end{equation}  
The $L^1$-completion of $C_c(\RR,\Bb)$, i.e.\ completion w.r.t.\ the
norm $\|f\|_1:=\int_{\RR}dx \,\|f(x)\|_{\Bb}$, is a
Banach algebra, the $L^1$-crossed product
denoted $L^1(\RR,\Bb,\alpha)$.
The crossed product algebra $\Bb\rtimes_\alpha\RR$
is the completion of $L^1(\RR,\Bb,\alpha)$ 
w.r.t.\ the $C^*$-norm  $\|f\|:=\sup_\rho
\|\rho(f)\|$ where the supremum is taken over all bounded
$*$-representations. It is not necessary to perform the middle step
via the $L^1$-crossed product, but it is sometimes convenient to 
work with it when verifying that the integral kernel of 
a given operator belongs to $\Bb\rtimes_\alpha\RR$. In this spirit, we can
benefit in Section \ref{obsalg} from our results in \cite{KS1}. 
By a continuity argument, one can simply work with functions $f:\RR\to
\Bb$ when performing calculations with elements of $\Bb\rtimes_\alpha
\RR$

\vspace{.2cm}

Let $(\rho,\H)$ be a representation of $\Bb$. 
It induces a
representation $(\pi,L^2(\RR,\H))$ of $\Bb\rtimes_{\alpha}\RR$:
\begin{equation}\label{rep} 
(\pi(f) \psi) (x) 
 \;=\;
 \int_{\RR} dy\,
 \rho(\alpha_{-x}(f(x-y))) \psi(y) 
 \mbox{ . }
\end{equation}

\subsection{$C^*$-fields}
\label{appendix2}

We follow the exposition of \cite{Landsman98} 
in defining a continuous field of \CA s 
(or simply a \CF)
$(\Cc,\{\Cc^\hbar,\varphi_\hbar\}_{\hbar\in I})$ over a locally
compact Hausdorff space $I$. This consists of a \CA\ $\Cc$ (also
called the total algebra of the field), a collection 
of \CA s $\{\Cc^\hbar\}_{\hbar\in I}$, one for each point of the space
$I$, with surjective algebra homomorphisms $\varphi_\hbar:\Cc\to
\Cc^\hbar$ such that, 
\begin{enumerate}
\item for $a\in\Cc$, $\|a\|=\sup_{\hbar\in
    I}\|\varphi_\hbar(a)\|$,
\item \label{cont}
for all $a\in\Cc$, $\hbar \mapsto \|\varphi_\hbar(a)\|$ is a
  function in $C_0(I)$,
\item $\Cc$ is a left $C_0(I)$ module and, for $f\in C_0(I), a\in \Cc$
  we have $\varphi_\hbar(fa) = f(\hbar) \varphi_\hbar(a)$.
\end{enumerate}
The construction is reminiscent of a fibre bundle, except there is no
typical fibre, the algebras $\Cc^\hbar$ need not to be isomorphic
even if $I$ is connected, and so one cannot define how the
$\Cc^\hbar$ are topologically glued together using local
trivializations. This information is contained in the algebra 
$\Cc$, the total \CA\ of the field. 
In fact, continuous sections of the field are
collections $\{a_\hbar\}_{\hbar\in I}$ for which 
exist $a\in\Cc$ such that $\varphi_\hbar(a)=a_\hbar$. 
$\Cc$ can therefore be seen as the algebra of continuous sections with
pointwise (in $\hbar$) multiplication. A \CF\ is called trivial if
$\Cc=C_0(I,\Bb)$ for some \CA\ $\Bb$, $\Cc^\hbar=\Bb$ 
and $\varphi_\hbar$ the evaluation at $\hbar$.

\vspace{.2cm}

All we are interested in here concerns the more special set up in which
$I\subset \RR$ and we have a collection of continuous 
$\RR$-actions $\{\alpha^\hbar\}_{\hbar\in I}$ on a single \CA\
$\Bb$, $\alpha^\hbar:\RR\to \mbox{\rm Aut}(\Bb)$.
Collecting these together we get an $\RR$ action
$\tilde\alpha:\RR\to \mbox{\rm Aut}(C_0(I,\Bb))$ by
$$\tilde\alpha_t(f)(\hbar)= \alpha_t^\hbar(f(\hbar))\;,$$ 
which is continuous provided the above expression is continuous in
$\hbar$ for all $t$ and $f$  which we hereby assume.
Then
$(C_0(I,\Bb)\rtimes_{\tilde\alpha}\RR,\{\Bb\rtimes_{\alpha^\hbar}\RR,
\eva_\hbar\}_{\hbar\in I})$ 
is a continuous field of \CA s \cite{Rieffel89}.

\begin{ex}[Heisenberg group algebra]\label{ex5a}
{\rm The (polarized) Heisenberg group
$\HH$ is $\RR^3$ as topological space, but with (non-abelian) multiplication 
$$(a_1,a_2,a_3)(b_1,b_2,b_3)=(a_1+b_1,a_2+b_2,a_3+b_3+a_1b_2)\;.
$$
It contains the subgroup $\RR^2=\{(a_1,a_2,a_3)\in \HH|a_1=0\}$
so that $\HH$ can be identified with the semi-direct product
$\RR^2\rtimes_{\tilde\tau}\RR$ where
$\tilde\tau_{a_1}(a_2,a_3)=(a_2,a_3+a_1a_2)$.
The Heisenberg group algebra (i.e.\ the crossed product
$\CC\rtimes_\aco\HH$ defined in a similar way as for $\RR$) can
therefore be identified with the
\CA\ $C_0(\RR^2)\rtimes_{\tilde\tau}\RR$ with 
$\tilde\tau_{a_1}(f)(a_2,a_3)=f(a_2,a_3-a_1a_2)$.
Let
$\varphi_{a_2}: C_0(\RR^2)\rtimes_{\tilde\tau}\RR
\to C_0(\RR)\rtimes_{\tau^{a_2}}\RR$ be evaluation of the
$2$-component at $a_2$, i.e.\ $\varphi_{a_2}(f)(a_1)(a_3)=
f(a_1)(a_2,a_3)$. Then 
$\im(\varphi_{a_2})\cong C_0(\RR)\rtimes_{\tau^{a_2}}\RR$ where
$\tau^{a_2}_{a_1}(g)(a_3)=g(a_3-a_2 a_1)$ for $g:\RR\to
C_0(\RR)$. Furthermore 
$(C_0(\RR^2)\rtimes_{\tilde\tau}\RR,\{C_0(\RR)\rtimes_{\tau^{a_2}}\RR,
\varphi_{a_2}\}_{a_2\in\RR})$ is
a $C^*$-field. Therefore $a_2$ plays the role of $\hbar$.
}\end{ex}

\begin{ex}\label{ex5} {\rm 
If we have an $\RR$-action $\alpha$ on a \CA\ $\Bb$ we can extend the
above field of the Heisenberg group algebra in the following way: With
the above $\RR$-action $\ttau$ on $C_0(\RR^2)$ define 
$\ttau\otimes\alpha:\RR\to{\rm Aut}\,C_0(\RR^2,\Bb)$ by
$(\ttau\otimes\alpha)_{a_1}(f)(a_2,a_3)=\alpha_{a_1}(f(a_2,a_3-a_2a_1))$. 
Setting $a_2=\hbar$ as above, this then yields a $C^*$-field 
$(C_0(\RR^2,\Bb)\rtimes_{\tilde\tau\otimes\alpha}\RR,
\{C_0(\RR,\Bb)\rtimes_{\tau^\hbar\otimes\alpha}\RR,  
\varphi_\hbar\}_{\hbar\in\RR})$ which will be of crucial importance 
later on.
This \CF\ is trivial away from $\hbar= 0$, i.e., for $\hbar\neq 0$,
$C_0(\RR,\Bb)\rtimes_{\tau^1\otimes\alpha}\RR\cong
C_0(\RR,\Bb)\rtimes_{\tau^\hbar\otimes\alpha}\RR$ and
$\ker( \varphi_0)
\cong C_0(\RR\backslash
\{0\},C_0(\RR,\Bb)\rtimes_{\tau^1\otimes\alpha}\RR)$. 
However, $C_0(\RR,\Bb)\rtimes_{\tau^0\otimes\alpha}\RR\cong 
C_0(\RR,\Bb\rtimes_{\alpha}\RR)$ is not isomorphic to
$C_0(\RR,\Bb)\rtimes_{\tau^1\otimes\alpha}\RR$. 
}\end{ex}

\subsection{Extensions}
\label{sec-extensions}

Suppose that we have a surjective morphism between \CA s 
$ q:\Cc {\to} \Bb$.  One then says that $\Cc$ is
an extension of $\Bb$ by the ideal $\Jj:=\ker(q)$.\footnote{Some 
authors call $ \Cc$ an extension of $\Jj$ by $\Bb$.} 

\begin{ex}[Cone of an algebra]\label{ex-cone} {\rm 
The suspension of the algebra $\Bb$ is
$S\Bb:=C_0(\RR,\Bb)$. Its cone is given by 
$C\Bb:=C_0(\RR\cup\{+\infty\},\Bb)$. The cone is an
extension of $\Bb$ by the ideal $S\Bb$, the morphism being
$q=\mbox{ev}_\infty$, the evaluation at $+\infty$.
}\end{ex}

\begin{ex}[Wiener-Hopf extension]\label{ex-WH} 
{\rm
Let $\alpha$ be an $\RR$-action on $\Bb$.
We extend the $\RR$-actions $\tau^\hbar\otimes\alpha$ on the suspension  
$S\Bb$ 
of Example~\ref{ex5} to the cone $C\Bb$ by setting
$(\tau^\hbar\otimes\alpha)_t f({+\infty}) = \alpha_t(f({+\infty}))$. 
Hence evaluation at ${+\infty}$ yields a surjective
algebra-homomorphism
\begin{equation}\label{WHext}
\eva_\infty:C\Bb\rtimes_{\tau^\hbar\otimes\alpha}\RR \longrightarrow
\Bb\rtimes_{\alpha}\RR\mbox{ . }
\end{equation} 
For $\hbar=1$ the corresponding extension is called the
Wiener-Hopf extension for an $\RR$-action
$\alpha$ on $\Bb$ \cite{Rieffel89}. The ideal is $\ker(\eva_\infty)=S\Bb
\rtimes_{\tau^1\otimes\alpha}\RR$ which appeared
in Example~\ref{ex5}, it is isomorphic to
$\K\otimes\Bb$ \cite{Rieffel82} (see also the appendix).
These form the ingredients of the exact sequence (\ref{eq-introext}). 
}\end{ex}

\begin{ex}[Extension of 
Heisenberg group algebra]\label{ex-Heisenextend} 
{\rm
By repeating the constructions of Example \ref{ex5} but with $\RR^2$
replaced by $\RR\times(\RR\cup\{+\infty\})$ and actions extended as above,
one obtains the 
$C^*$-field 
$(C_0(\RR,C\Bb)\rtimes_{\tilde\tau\otimes\alpha}\RR,
\{C\Bb\rtimes_{\tau^\hbar\otimes\alpha}\RR,  
\varphi_\hbar\}_{\hbar\in \RR})$. 
The map in (\ref{WHext}) now extends to a surjection which we also denote by 
$\eva_\infty$, 
\begin{equation}\label{WHextext}
\eva_\infty : C_0(\RR,C\Bb)\rtimes_{\tilde\tau\otimes\alpha}\RR 
\to C_0(\RR,\Bb)\rtimes_{\alpha}\RR\mbox{ , }
\end{equation}
whose kernel is $C_0(\RR,S\Bb) \rtimes_{\tilde\tau\otimes\alpha}\RR$.
Each algebra is the total algebra of a \CF\ so that one actually has a
field of extensions, the fibre at $\hbar=1$ being the Wiener-Hopf extension.
}\end{ex}

\section{$K$-theoretic preliminaries}
\label{sec-K}

This introduction is mainly meant to fix notations.
For a complete definition of the $K$-groups for a Banach 
algebra $\Bb$, {\sl cf.} \cite{Bla}. We denote by $[\Bb]_0$ the homotopy
classes of projections of $\Bb$ and by $\Bb^+$ the unitalisation,
$\Bb^+=\Bb\times\CC$ with
$(A,\lambda)(A',\lambda')=(AA'+\lambda A'+A\lambda',\lambda\lambda')$. 
The $C^*$-inductive limit of the matrix algebras $M_n(\Bb^+)$
is denoted by $M_\infty(\Bb^+)$. 
$[M_\infty(\Bb^+)]_0$ is a monoid under 
addition of homotopy classes of projections, $[p]_0+[q]_0=[\diag{p}{q}]_0$. 
The $K_0$-group $K_0(\Bb)$
of $\Bb$ is obtained from the monoid $[M_\infty(\Bb^+)]_0$ by
Grothendieck's construction and then factorizing out the added unit. 

\vspace{.2cm}

Let $U(\Bb)$
be the group of unitaries $u\in \Bb^+$ such that $u-1\in \Bb$ (the
$1$ is here the unit in $\Bb^+$). We denote by $[\Bb]_1$ the homotopy
classes of $U(\Bb)$. 
The algebraic limit of the groups $U(M_n(\Bb))$ is denoted by
$U(M_\infty(\Bb))$ and then $K_1(\Bb)=[M_\infty(\Bb)]_1$.
The (non-abelian) product in
$U(M_n(\Bb))$ induces a product in $[M_\infty(\Bb)]_1$ which is abelian and
therefore denoted additively.

\subsection{Elliott-Natsume-Nest map}

Suppose we have a continuous field of \CA s
$(\Cc,\{\Cc^\hbar,\varphi_\hbar\}_{\hbar\in I})$ over $I=[0,1]$ 
which is trivial away from $\hbar=0$. This means that there are
isomorphisms $\phi_\hbar:\Cc^1\to \Cc^\hbar$ for $\hbar>0$ such that
$\phi:C_0((0,1],\Cc^1)\to \ker(\varphi_0)$: 
$\varphi_\hbar(\phi(f))= \phi_\hbar(f(\hbar))$ is an
isomorphism. The following theorem shows that in this situation one
obtains maps $[\Cc^0]_i\to [\Cc^1]_i$ which induce homomorphisms 
$K_i(\Cc^0)\to K_i(\Cc^1)$. We call these maps ENN-maps.

\begin{thm}\label{thmENN} 
{\rm \cite{ENN2}} Consider a continuous field of \CA s
$(\Cc,\{\Cc^\hbar,\varphi_\hbar\}_{\hbar\in I})$ over $I=[0,1]$ 
which is trivial away from $\hbar=0$. For any projection 
$p\in\Cc^0$ there is a projection valued section $\tilde p \in \Cc$ such that
$\varphi_0(\tilde p) = p$. For any $u\in U(\Cc^0)$ there is a 
section $\tilde u \in U(\Cc)$ such that
$\varphi_0(\tilde u) = u$.
The maps $\enn_i:[\Cc^0]_i\to [\Cc^1]_i$:
$\enn_0([p]_0)=[\varphi_1(\tilde p)]_0$, 
$\enn_1([u]_1)=[\varphi_1(\tilde u)]_1$
are well-defined and induce homomorphisms 
$\enn_i:K_i(\Cc^0)\to K_i(\Cc^1)$.
\end{thm}

\noindent
\bew\ (We only recall how these maps are constructed, for the rest see
\cite{ENN2}.) 
Let $p$ be a projection in $\Cc^0$. Since $\varphi_0$ is
surjective, there exists a selfadjoint section 
$x\in\Cc$ with $\varphi_0(x)=p$. By property \ref{cont}.\ of \CF s,
we find for any $\delta>0$ an $\epsilon$ such that
$\|\varphi_\hbar(x^2-x)\|<\delta$ for $\hbar<\epsilon $.
For small $\delta$ the spectrum of $\varphi_\hbar(x)$ is close to
$\{0,1\}$ and we can find a continuous function $f:\RR\to \RR$, 
vanishing for $t<a$ and being $1$ for $t>b$ where $0<a<b<1$ and
$(a,b)$ does not intersect the spectra of $\varphi_\hbar(x)$,
$\hbar\leq\epsilon$. 
Then $f(x)$ is another section with   
$\varphi_0(f(x))=p$, but such that $\varphi_\hbar(f(x))$ are
projections for $\hbar\leq\epsilon $. Now the section can be extended
by the constant section
since the field is trivial away from $\hbar=0$. The resulting
section is $\tilde p$ where 
$\varphi_\hbar(\tilde p)=\varphi_\epsilon(f(x))$ if $\hbar\geq\epsilon$.

\vspace{.2cm}

The choice of $x$ is not canonical, 
but it is not difficult to see that the homotopy
class of $\tilde p$ is uniquely determined since any other choice
$\tilde p'$ needs to be close to $\tilde p$ at small $\hbar$. 
The case of unitaries works in a similar way.

\vspace{.2cm}

With canonically extended
$\varphi_\hbar$, the field
$(M_n(\Cc^+),\{M_n({\Cc^\hbar}^+),
\varphi_\hbar\}_{\hbar\in I})$ 
is a continuous field of \CA s which is trivial away
from $0$. The above construction applies therefore also to elements in
$[M_n({\Cc^0}^+)]_0$ and $[M_n({\Cc^0})]_1$
and induces homomorphisms between the corresponding $K$-groups. 
\eb

\subsection{Boundary maps in $K$-theory}\label{sec-exp}

Suppose given an extension $ \Cc \stackrel{q}{\to} \Bb$
by $\Jj:=\ker(q)$.
What interests us here are two maps, the boundary maps in $K$-theory,
which measure the
extend to which the map induced by $q$ on homotopy classes is not
surjective.
The first of these maps is the exponential map
$$\exp\, :\, K_0(\Bb)\to K_1(\Jj)$$
which is induced from the map
$\exp: [\Bb]_0\to [\Jj]_1$  defined as follows: 
Let $p$ be a projection in
$\Bb$.  Since $q$ is surjective, there
exists an $x\in \Cc$ such that $q(x)=p$ . Since $p$ is selfadjoint
we can choose $x$ selfadjoint and define
$$\exp[p]_0\;:=\;[u]_1\;,\qquad u \;=\; e^{2\pi \imath x}\;.$$

If we apply the above to the cone (Example~\ref{ex-cone})
given by  $C\Bb\stackrel{\eval_\infty}{\to}\Bb$, 
then $\ker( \eva_\infty)$ is the suspension of
$\Bb$ and the exponential map is the so-called Bott
map $\exp=\beta:K_0(\Bb)\to K_1(S\Bb)$,
$$\beta[p]_0 = [e^{2\pi \imath \chi p}]_1$$
where $\chi:\RR\to [0,1]$ is a continuous function with
$\lim_{t\to-\infty}\chi(t)=0$ and $\lim_{t\to\infty}\chi(t)=1$. 

\vspace{.2cm}

The second map of interest is the index map
$\mbox{ind} : K_1(\Bb)\to K_0(\Jj)$ defined as follows:
Given $V\in U(M_n(\Bb))$ defining a class in $K_1(\Bb)$, let $W\in
U(M_{2n}(\Bb))$ be a lift of
$\left( \begin{array}{cc} V & 0 \\ 0 & V^*\end{array}\right)$. Then
$$
\mbox{ind}([V]_1)\;=\;
\left[W\left(\begin{array}{cc} 1 & 0 \\ 0 & 0 \end{array}
\right)W^*\right]_0
-
\left[\left(\begin{array}{cc} 1 & 0 \\ 0 & 0 \end{array}
\right)
\right]_0
\mbox{ . }
$$
The index map of the extension defined by
$C\Bb\stackrel{\eval_\infty}{\to}\Bb$
is denoted by $\Theta$. The fact \cite{Bla} that the
compositions $\Theta\beta:K_0(\Bb)\to
K_0(SS\Bb)$ and $\beta\Theta:K_1(\Bb)\to
K_1(SS\Bb)$ are isomorphisms is called Bott periodicity.  

\subsection{Boundary maps of Wiener-Hopf extension}

We want to express 
the exponential and the index map of the Wiener-Hopf extension 
$\eva_\infty:C\Bb\rtimes_{\tau^1\otimes\alpha}\RR \longrightarrow
\Bb\rtimes_{\alpha}\RR$ (discussed in Example \ref{ex-WH})
using an ENN-map. 
Herefore we use the $C^*$-fields of Example \ref{ex-Heisenextend} 
restricted to $[0,1]\in\RR$. They form the extension
\begin{equation}
\label{ses}
\eva_\infty : C([0,1],C\Bb)\rtimes_{\tilde\tau\otimes\alpha}\RR 
\to
C([0,1],\Bb)\rtimes_{\alpha}\RR\;.
\end{equation}
The \CF\ 
corresponding to $C([0,1],\Bb)\rtimes_{\alpha}\RR$ is trivial
and that corresponding to the ideal
$C([0,1],S\Bb) \rtimes_{\tilde\tau\otimes\alpha}\RR$
satisfies the conditions of Theorem~\ref{thmENN} to give rise to
ENN-maps 
\begin{equation}\label{ENN-map}
\enn_i:[S\Bb \rtimes_{\aco\otimes\alpha}\RR]_i\to 
[S\Bb \rtimes_{\tau\otimes\alpha}\RR]_i\;.
\end{equation}

\begin{prop} 
\label{prop-ENNWH}
Let $\exp$ and $\mbox{\rm ind}$ be exponential and index
  map of the Wiener-Hopf extension {\rm (\ref{WHext})}. Then
$\enn_1\beta = \exp$ and $\enn_0\Theta = \mbox{\rm ind}$. 
Here we have used the identification $C\Bb
\rtimes_{\aco\otimes\alpha}\RR\cong C(\Bb \rtimes_{\alpha}\RR)$.
\end{prop}

\noindent 
\bew\ A projection $p\in \Bb\rtimes_\alpha\RR$ defines a constant section in 
$C([0,1],\Bb)\rtimes_{\alpha}\RR$. If $x\in
C([0,1],C\Bb)\rtimes_{\tilde\tau\otimes\alpha}\RR$ is a selfadjoint
lift of the constant section 
under (\ref{ses}) then, by definition, $\enn_1[e^{2\pi i
  \varphi_0(x)}]_1 = [e^{2\pi \imath \varphi_1(x)}]_1$. Furthermore 
$\exp[p]_0=[e^{2\pi \imath \varphi_1(x)}]_1$ since
$\varphi_1(x)$ is a lift of $p$ in (\ref{WHext}). The claim
follows since $\varphi_0(x)$ is a lift of $p$ in the extension 
$C\Bb\rtimes_{\aco\otimes\alpha}\RR 
\;\stackrel{\varphi_\infty}{\to}\;
\Bb\rtimes_{\alpha}\RR$, and $\chi p$  a lift of $p$ in the extension 
$C(\Bb\rtimes_{\alpha}\RR) 
\;\stackrel{\varphi_\infty}{\to}\;
\Bb\rtimes_{\alpha}\RR$. Under the identification stated in the lemma,
$e^{2\pi \imath \chi p}$ is therefore homotopic to  $e^{2\pi 
\imath \varphi_0(x)}$.
The argument involving the index map is similar.
\eb

\section{Higher traces on Banach algebras}
\label{sec-cycliccoh}

For background information on cyclic cohomology and higher traces (or
$n$-traces) see
\cite{ConnesBook}.
Given an associative algebra $\Bb$ 
let $C_\lambda^n(\Bb)$ be the set of $n+1$-linear functionals on $\Bb$
which are cyclic in the sense that
$\eta(A_1,\cdots,A_n,A_0) = (-1)^n \eta (A_0,\cdots,A_n)$.
\noindent 
Define the boundary operator $b:C_\lambda^n(\Bb)\to C_\lambda^{n+1}(\Bb)$:
$$
b\eta (A_0,\cdots,A_{n+1})\;=\; 
\sum_{j=0}^n (-1)^j \eta(A_0,\cdots,A_jA_{j+1},\cdots ,A_{n+1})
+ (-1)^{n+1} \eta(A_{n+1}A_0,\cdots,A_{n})\;.
$$
An element $\eta \in C^n_\lambda (\Bb)$ satisfying
$b\eta=0$ is called a cyclic $n$-cocycle and
the cyclic cohomology $HC(\Bb)$ of $\Bb$ is the cohomology of the
complex
$
0\to C^0_\lambda (\Bb)\to
\cdots\to C^n_\lambda (\Bb)
\stackrel{b}{\to} C^{n+1}_\lambda (\Bb) 
\to\cdots$.


\subsection{Cycles}
\label{sec-cycles}

A very convenient way of
looking at cyclic cocycles is in terms of characters of graded differential 
algebras with graded closed traces $(\Omega,d,\int)$ over $\Bb$.
Here $\Omega=\bigoplus_{n\in\Nat_0}\Omega^n$ is a graded algebra 
(we denote by $\deg(a)$ the degree of a homogeneous element $a$)
and $d$ is a graded differential on $\Omega$
of degree $1$. A graded trace on the subspace
$\Omega^n$ is a linear functional $\int:\Omega^n\to \Complex$ which 
is cyclic in the sense that
$\int w_1w_2 = (-1)^{\deg(w_1)\deg (w_2)}\int w_2 w_1$. It is closed
if it vanishes on $d(\Omega^{n-1})$. In the situation below there is a
largest number $n$ for which $\Omega^n$ is non-trivial. This $n$ 
is called the top degree of $\Omega$. The
graded trace will be a graded trace on the sub-space of 
top degree.

\begin{defn}\label{cycle}
An $n$-dimensional cycle is a
graded differential algebra $(\Omega,d)$
of top degree $n$
together with a closed graded trace $\int$ on $\Omega^n$.
A cycle $(\Omega,d,\int)$ is called a cycle over $\Bb$ if there is
an algebra homomorphism
$\Bb\to \Omega^0$.
\end{defn}

We will assume here that the homomorphism
$\Bb\to \Omega^0$ is injective and hence
identify $\Bb$ with a sub-algebra of $\Omega^0$.
The connection with cyclic cocycles is given by the following
proposition \cite{ConnesBook}. 

\begin{prop}
Any cycle of dimension $n$ over $\Bb$ defines a cyclic $n$-cocycle
through what is called its character:
$$
\eta(A_0,\ldots,A_n) 
\;= \;
\int A_0 d A_1\cdots d A_n 
\;.
$$
Conversely, any cyclic $n$-cocycle arises as the character of an $n$-cycle.
\end{prop}

A (bounded) trace over $\Bb$ is an example of a cyclic
$0$-cocycle. Taking
$\Omega=\Bb$, $d=0$, $\int$ to be that trace, we have a realization of
the trace as character of a $0$-cycle.  

\vspace{.2cm}

For our purposes, the cyclic cohomology of $C^*$-algebras
is too small, because we need multilinear functionals which are unbounded.
A particular class of unbounded cyclic cocycles suitable for
our purposes is given by the higher traces
\cite{ConnesBook,Connes86}. 
These are
characters of cycles over dense sub-algebras $\Bb'$ of $\Bb$
satisfying a continuity condition. 
It will be useful to relax the
requirement of $\Bb$ being a \CA\ and rather consider Banach
algebras.

\begin{defn} \label{def3}
An $n$-trace on a Banach algebra $\Bb$ is the character of
an $n$-cycle
 $(\Omega',d,\int)$ over a dense sub-algebra $\Bb'$ of $\Bb$ such that
for all $A_1,\ldots, A_n\in\Bb'$ there exists a constant $C=C(A_1,\dots,A_n)$ 
such that 

\begin{equation}\label{topcond}
\left|\int (X_1dA_1)\cdots  (X_n dA_n)\right| 
\;\leq\; 
C\|X_1\|\cdots\|X_n\|
\;,
\end{equation}

\noindent for all $X_j\in {\Bb'}^+$.
\end{defn}

Condition (\ref{topcond}) may be rephrased by saying that for all
$A_1,\ldots, A_n\in\Bb'$ the apriori densely defined
multi-linear functional

$$
\Bb^{\times n}\to \CC\,:\quad 
(X_1,\dots,X_n)\; \mapsto\;
\int (X_1dA_1)\cdots  (X_n dA_n)
$$

\noindent extends to a bounded multi-linear functional. 
Denoting by $p(A_1,\dots,A_n)$ the norm of that
functional, i.e.\ the best possible constant $C$ in (\ref{topcond}),
we have a family of maps  $\Bb^{\times n}\to \RR$ satisfying 
$$ 
p(A_1,\dots,\lambda A_j+\lambda' A_j',\dots,A_n)
\;\leq\;
|\lambda|\,p(A_1,\dots,A_j,\dots,A_n)+
|\lambda'| \,p(A_1,\dots,A_j',\dots,A_n)
\;.
$$
\noindent But since $d$ is a derivation, it also satisfies
$$ 
p(A_1,\dots,A_jA_j',\dots,A_n)
\;\leq\;
\|A_j'\|\,p(A_1,\dots,A_j,\dots,A_n)+
\|A_j\|\, p(A_1,\dots,A_j',\dots,A_n)
\;
.
$$

For simplicity, rather than considering cycles $(\Omega',d,\int)$
over a dense sub-algebra $\Bb'$,  we shall
consider triples $(\Omega,d,\int)$ as in Definition~\ref{cycle} 
with $\Omega$ being a Banach algebra,
$\Bb\subset \Omega^0$, but allowing for the
possibility that $d$ and $\int$ are only densely defined. 
If the character is densely
defined and satisfies (\ref{topcond}), we call the triple
$(\Omega,d,\int)$ an unbounded $n$-cycle. The role of (\ref{topcond})
is to insure the existence of a third algebra $\Bb''$,
$\Bb'\subset\Bb''\subset \Bb$, to which the character can be extended
(by continuity) and such that the inclusion $i:\Bb''\hookrightarrow
\Bb$ induces an isomorphism between $K(\Bb'')$ and $K(\Bb)$ 
\cite{Connes86}. 

\vspace{.2cm}

An example of a cycle for the commutative algebra $\Bb=C(M)$ of 
continuous functions over a compact manifold without boundary
is given by the algebra of exterior forms with its usual differential
$(\Omega(M),d)$ and graded trace equal to integration of
$n$-forms, $n=\dim (M)$. This is an unbounded cycle. One may take 
$\Bb'=C^\infty(M)$ and $p(A_1,\dots,A_n)=\int |dA_1\cdots dA_n|$
where (locally) $|dA_1\cdots dA_n|=|f|d\,\mbox{\rm vol}$ if 
$dA_1\cdots dA_n=fd\,\mbox{\rm vol}$. Note that $p(A_1,\dots,A_n)$ is
not continuous in $A_j$ w.r.t.\ the supremum norm, which is the
$C^*$-norm of $\Bb$.

\vspace{.2cm}

A $0$-trace is a (possibly unbounded) 
linear functional $\tr$ which is cyclic and satisfies
(\ref{topcond}). A positive trace is a positive linear
functional $\tr$ which is cyclic. It might be unbounded (with
dense domain), but it always satisfies
$|\tr(AX)|\leq \tr(|A|)\|X\|$ if $A$ is trace class and hence 
(\ref{topcond}) holds with $\Bb'$ being the ideal of trace class elements.

\vspace{.2cm}

Here we need to construct higher traces on a Banach algebra $\Bb$
on which is given a differentiable action of $\Real^n$ leaving 
a (possibly unbounded) trace invariant. This is essentially
Ex.~12, p.~254 of \cite{ConnesBook}.

\begin{prop}\label{prop1a}
Let $\Bb$ be a Banach 
algebra with a differentiable action of $\Real^n$ and
$\TV$ be an invariant positive trace on $\Bb$. 
Denote by 
$\nabla_j$, $j=1,\cdots,n$, commuting closed derivations defined by
the action and suppose that 
$\Bb' = \{
A\in\bigcap_{j=1}^n 
\dom(\nabla_j)| \,\exists j:\nabla_j A\mbox{ traceclass}\}
$
is dense in $\Bb$. 
Then $(\Omega,d,\int)$ is an unbounded $n$-cycle over $\Bb$ where
$$
\Omega\;:=
\;\Bb\otimes\Lambda \CC^n\;,
$$ 
the tensor product of $\Bb$ with the Grassmann algebra $\Lambda \CC^n$ 
with generators $e_j$, $j=1,\dots,n$,
$$
d(A\otimes v)
\;=\;
\sum_{j=1}^n\nabla_j A \otimes e_j v\;,
$$
and 
$\int = \TV\otimes\imath$ with $\imath(e_1\cdots e_n)=1$, explicitly 
$$
\int A_0d A_1\cdots d A_n
=\sum_{\sigma\in S_n}\mbox{\rm sgn}(\sigma)\,
 \TV(A_0\nabla_{\sigma(1)}A_1\cdots\nabla_{\sigma(n)}A_n)\; .
$$
\end{prop}

\noindent
\bew\ The algebraic aspects of this proposition are straightforward to show, 
see e.g.\ \cite{KRS02}. Since trace class operators form an ideal, 
$\Bb'$ is a sub-algebra. Then 
(\ref{topcond}) follows from 
$$
|\TV((X_1\nabla_{1}A_1)\cdots(X_n\nabla_{n}A_n))|
\;\leq\; 
\|X_1\|\cdots \|X_n\| \,\|\nabla_{1}A_1\|\cdots
\|\nabla_{n-1}A_{n-1}\|
\; \TV(|\nabla_{n}A_n|)\;,
$$
and the cyclicity of $\TV$.\eb

\vspace{.2cm}

The following is an extension of the above construction, it
corresponds to an iteration of Lemma~16 p.~258 of \cite{ConnesBook}.

\begin{prop}
\label{prop1}
Let $(\Omega,d,\int)$ be a {\rm (}possibly unbounded{\rm )} $k$-cycle over 
the Banach algebra $\Bb$ which is invariant under a differentiable action of 
$\Real^n$ in the sense that this action commutes with $d$ and
leaves $\int$ invariant. 
Denote by 
$\nabla_j$, $j=1,\cdots,n$, commuting closed derivations defined by
the action and suppose that $\bigcap_{j=1}^n \dom(\nabla_j) \cap \Bb'$ 
is a dense sub-algebra of $\Bb$ such that on $\Bb'\subset \Bb$ 
the character of $(\Omega,d,\int)$ is fully defined. 
Taking $\Omega'=\Omega\hat\otimes \Lambda\CC^n$, the
graded tensor product,
$d'=d\hat\otimes 1+\delta$ with
$\delta(w\hat{\otimes} v) =  
(-1)^{\partial w}\sum_j \nabla_j w\hat\otimes e_j v$
and $\int' = \int\hat{\otimes}\,\imath$,
one obtains a $k+n$-cycle $(\Omega',d',\int')$ over $\Bb$.
\end{prop}

\noindent {\bf Proof:}
The algebraic aspects are straightforward and again given in
 \cite{KRS02}. The only point to settle is condition
(\ref{topcond}). It follows iteratively from the case $n=1$. For
$n=1$, using cyclicity,
\begin{eqnarray*}\left|\int'(X_1d'A_1)\dots(X_{k+1}d'A_{k+1})\right| &\leq& 
\sum_{j=1}^{k+1}\left| \int (X_j\nabla_1
  A_j)(X_{j+1}\delta A_{j+1})\cdots(X_{j-1}\delta A_{j-1})\right|\\
&\leq& 
\sum_{j=1}^{k+1}\|X_1\|\cdots \|X_{k+1}\|\,
\left\|\nabla_1 A_j\right\|
  C_j,
\end{eqnarray*}
where $C_j$ depends only on $A_1,\dots,A_{j-1},A_{j+1},\dots,A_{k+1}$.
This inequality shows also that the character of the cycle is defined
on $\bigcap_{j=1}^n \dom(\nabla_j) \cap \Bb'$.\eb 

\subsection{Cyclic cocycles for crossed products with $\RR$}
\label{sec-cross}

An action of $\RR$ on a graded 
differential algebra $(\Omega,d)$ is a
homomorphism $\alpha:\RR\to \mbox{\rm Aut}(\Omega)$ such that $\forall\;
t\in\RR$, $\alpha_t$ has degree $0$ and commutes with $d$. 
If $\Omega$ is a Banach algebra or even a \CA, 
we require in addition that 
for all $A\in\Bb$, $t\mapsto
\alpha_t(A)$ is continuous and $\|\alpha_t\|=1$. Therefore
we can form 
$L^1(\Omega,\RR,\alpha)$ as well as
the crossed product $\Omega\cp_\alpha\RR$. 

\begin{defn}\label{def4}
A $n$-cycle $(\Omega,d,\int)$ over $\Bb$
is called invariant under an action 
$\alpha$ of $\,\RR$ on $\Omega$ if the graded trace $\int$ is invariant 
under it. If $(\Omega,d,\int)$ is unbounded, we require in addition
that the norms $p(A_1,\dots,A_n)$ {\rm ({\sl cf.}\ Definition~\ref{def3})}
satisfy that
\begin{equation}
\label{eq-ntracebound} 
Q(A_1,\dots,A_n)
\;:=\;
\sup_{t_i\in \RR}\,
p(\alpha_{t_1}(A_1),\dots,\alpha_{t_n}(A_n))
\end{equation}
is finite for all $A_j\in \Bb'\subset \Bb$ where $\Bb'$ is 
a dense sub-algebra on which
the character of the $n$-cycle is fully defined.
An $n$-trace of $\Bb$ is invariant under an action 
$\alpha$ of $\RR$ if it is the character of an $\alpha$-invariant cycle
$(\Omega,d,\int)$. 
\end{defn}

We note that, by  cyclicity of the graded trace, the above
additional condition is equivalent to demanding that  $\sup_{t_1\in \RR}
p(\alpha_{t_1}(A_1),A_2,\dots,A_n)$ exists for all $A_j\in \Bb'$.
Furthermore, $Q$ inherits the properties of $p$, i.e.\
\begin{equation}\label{semi1} 
Q(A_1,\dots,\lambda A_j+\lambda' A_j',\dots,A_n)
\;\leq\;
|\lambda|\,Q(A_1,\dots,A_j,\dots,A_n)+
|\lambda'| \,Q(A_1,\dots,A_j',\dots,A_n)\,,
\end{equation}
\begin{equation}\label{semi2} 
Q(A_1,\dots,A_jA_j',\dots,A_n)
\;\leq\;
\|A_j'\|\,Q(A_1,\dots,A_j,\dots,A_n)+
\|A_j\|\, Q(A_1,\dots,A_j',\dots,A_n)\,.
\end{equation}
\begin{thm}\label{prop2}
Let $(\Omega,d,\int)$ be an $\alpha$-invariant {\rm (}possibly
unbounded{\rm )} 
$n$-cycle over the Banach algebra $\Bb$ and 
$\nabla:L^1(\RR,\Bb,\alpha) \to L^1(\RR,\Bb,\alpha)$ 
be the derivation 
$\nabla f(x) = \imath x f(x)$. 
Then $(\Omega_\alpha,d_\alpha,\int_\alpha)$
is an unbounded $n+1$-cycle over 
$L^1(\RR,\Bb,\alpha)$
where
$$\Omega_\alpha = L^1(\RR,\Omega,\alpha)\, \hat\otimes\, \Lambda\Complex,$$
$$d_\alpha(\omega\hat\otimes v) 
\;=\;
 d'\omega\hat\otimes v + (-1)^{\deg(\omega)}
\,\nabla \omega\hat\otimes \,e_1 v\;,
\qquad 
d'\omega(x) \;= \;d(\omega(x))\;,\quad
\nabla\omega(x)\; =\; \imath x \omega(x)\;,$$
and $\int_\alpha=  
\int\,\mbox{\rm ev}_0\hat\otimes\;\imath$, 
i.e.\
$$
\int_\alpha f_0d_\alpha f_1\cdots d_\alpha f_{n+1}
\;=\;
\sum_{j=1}^{n+1}
(-1)^{n+1-j}\int
\left(f_0 d f_1\cdots
d f_{j-1} (\nabla f_j) d f_{j+1}\cdots df_{n+1}\right)(0).
$$
\end{thm}

\noindent 
\bew\ 
We first show that the triple $(L^1(\RR,\Omega,\alpha),d',\int\eva_0)$
defines an unbounded $n$-cycle over $L^1(\RR,\Bb,\alpha)$. 
The required algebraic properties are straightforwardly checked ({\sl cf.}
\cite{KRS02}) and we
focus here on the continuity aspects (\ref{topcond}).  
Let $\Bb'\subset\Bb$ be a dense sub-algebra on which the
character of $(\Omega,d,\int)$ is fully defined and
$${\mathcal V}^{\fin}
\;:=\;
\bigcup_{V\subset\Bb',\,\dim V<\infty} C_c(\RR,\overline{V})
\mbox{ , }
$$
the union being over all finite dimensional linear sub-spaces of
$\Bb'$ and
$\overline{V}:= \bigcup_{t\in \RR}\alpha_t(V)$, the orbit of $V$
under the action. The space
${\mathcal V}^{\fin}$ is linear and since $Q(A_1,\dots,A_n)$ is finite
for all $A_j\in \Bb'$, we obtain from (\ref{semi1}) 
that 
$$\overline{Q}(f_1,\dots,f_n)
\;:=\;
\sup_{t_j\in \RR} Q(f_1(t_1),\dots,f_n(t_n))$$ 
is finite for all
$f_j\in {\mathcal V}^{\fin}$. Furthermore, ${\mathcal V}^{\fin}$ is
clearly dense in  
$C_c(\RR,\Bb')$ in the $L^1$-norm. Now let
${\mathcal A}^{\fin}$ be the sub-algebra of $L^1(\RR,\Bb,\alpha)$ 
generated algebraically by ${\mathcal V}^{\fin}$, {\sl i.e.}
it consists of finite twisted convolution products of elements of
${\mathcal V}^{\fin}$. Since, by (\ref{semi2}),
$$
Q(f_1g_1(t),f_2(t)\dots)
\;\leq\;
\int ds\Big(
\|f_1(s)\|\, 
Q(g_1(t-s),f_2(t)\dots)
\;+\;
\|g_1(t-s)\| Q(f_1(s),f_2(t)\dots)
\Big)
\;,
$$
we obtain
$$
\overline{Q}(f_1g_1,f_2\dots,f_n)
\;\leq\;  
\|f_1\|_{L^1} \,
\overline{Q}(g_1,f_2\dots)+\|g_1\|_{L^1}\, 
\overline{Q}(f_1,f_2\dots)
\,.
$$
Therefore $\overline{Q}(f_1,\dots,f_n)$ is finite for all $f_j\in {\mathcal
  A}^{\fin}$. 
The character of
$(L^1(\RR,\Omega,\alpha),d',\int\eva_0)$ is now restricted 
to the dense sub-algebra
${\mathcal A}^{\fin}$. Then 
we obtain
\begin{equation}\label{ineq1}
\left| \int \, \eva_0 (X_1d'f_1)\dots (X_nd'f_n)\right| 
\;\leq\; 
\|X_1\|_{L^1}\cdots \|X_n\|_{L^1} 
|\supp(f_1)|\cdots |\supp(f_n)|
\;\overline{Q}(f_1,\dots,f_n)\, .
\end{equation}
Here $|\supp(f)|$ is the (finite) length of the support of $f$. 
This shows that there exists a dense sub-algebra of
the Banach algebra $L^1(\RR,\Bb,\alpha)$ on
which the character of $(L^1(\RR,\Omega,\alpha),d',\int\, \eva_0)$ 
satisfies (\ref{topcond}). 

\vspace{.2cm}

Now on $L^1(\RR,\Bb,\alpha)$ and $L^1(\RR,\Omega,\alpha)$
we have the dual action of $\RR$ (identified here with the dual group
of $\RR$) and its corresponding
derivation is $\nabla$. It is densely defined and satisfies the conditions 
of Proposition~\ref{prop1}. Applying that proposition, one obtains
$(L^1(\RR,\Omega,\alpha),d_\alpha,\int_\alpha)$, which is 
an unbounded $n+1$-cycle.
\eb

\begin{defn}
If $\eta$ is the character of an $\alpha$-invariant $n$-cycle
$(\Omega,d,\int)$ over $\Bb$, we define
$\#_{\alpha}\eta$ to be
the character of $(\Omega_\alpha,d_\alpha,\int_\alpha)$ constructed in
{\rm Theorem~\ref{prop2}}, i.e.\ for $f_j\in L^1(\RR,\Bb,\alpha)$ 
$$
\#_\alpha\eta(f_0,\ldots,f_{n+1})
\;=\;
\sum_{k=1}^{n+1}\;(-1)^k\;\int \eva_0
\left(f_0d f_1\cdots\nabla f_k\cdots df_{n+1}\right)
\;, \qquad \nabla f(x)=\imath x f(x)\;.
$$
\end{defn}

Restricted to 
the context of smooth $\RR$-actions on smooth sub-algebras of \CA s,
Theorem~\ref{prop2} can be compared with 
a result in \cite{ENN1}. In fact, it coincides
with a construction given in that article for general
$n$-cycles over a smooth sub-algebra of $\Bb$ 
in the case that these $n$-cycles are $\alpha$-invariant in our sense.

\vspace{.2cm}

Whenever we have several commuting $\RR$ actions 
leaving a cycle invariant,
we can iterate this
construction, since we can extend a second action $\beta$ in $\Omega$ to an
action on $\Omega_\alpha$ commuting with the differential $d_\alpha$
and leaving $\int_\alpha$ invariant by evaluating it pointwise on
functions $f:\RR\to \Omega$ and keeping the new
Grassmann generator of $\Omega_\alpha$ fixed.

\begin{ex}[Suspension of $n$-traces] \label{prop3} {\rm 
The suspension $S\Bb$ of a \CA\ $\Bb$ is via Fourier transform isomorphic
to the crossed product $\Bb\rtimes_\aco\RR$.
For a given $n$-trace $\eta$ over $\Bb$, 
the above construction yields an $n+1$-trace $\#_\aco \eta$ over 
$L^1(\RR,\Bb,\id)$. When
intertwined with the Fourier transform one obtains  
a $n+1$-trace which we denote by $\eta^s$ over a dense Banach
sub-algebra of $S\Bb$.
However,
since the conditions of Definition~\ref{def4} are trivially
satisfied in that case
and the linear space ${\mathcal V}^{\fin}$ used in the proof
of Theorem~\ref{prop2}
is a dense sub-algebra of
$C_0(\RR,\Bb)$ under pointwise multiplication we can simplify 
the arguments of
Theorem~\ref{prop2} thereby improving (\ref{ineq1}),
namely, for $f_i\in {\mathcal V}^{\fin}$,
\begin{equation}\label{ineq1b}
\left| \int_\RR ds\int (X_1d'f_1)(s)\dots (X_nd'f_n)(s)\right| 
\;\leq\; 
\|X_1\| \cdots \|X_n\| \;\tilde {Q}(f_1,\dots,f_n)
\end{equation}
where $\tilde {Q}(f_1,\dots,f_n)=
|\supp(f_1)|\cdots
|\supp(f_n)|\,\sup_{t_j\in\RR}p(f_1(t_1),\dots,f_n(t_n))$ 
and the norm is here the $C^*$-norm on $S\Bb$.

\vspace{.2cm}

As a result, if $(\Omega,d,\int)$ is an $n$-cycle over a Banach
algebra $\Bb$ whose character is $\eta$ 
and $\partial_s:S\Bb\to S\Bb$ the derivative w.r.t.\ the suspension
variable, then
$\eta^s$ is the character of the unbounded $n+1$-cycle
$(\tOmega,\td,\tint)$ over $S\Bb$  where
$\tOmega:=S\Omega\hat\otimes\Lambda\Complex$
and ($\omega\in S\Omega$)
$$\td(\omega\hat\otimes v) 
\;=\; 
d'\omega\hat\otimes\, v\; +\; (-1)^{\deg(\omega)}\,
\partial_s\omega\hat\otimes \,e_1 v\;,$$
where $  d'\omega(s) = d(\omega(s))$
and $\tint=\int_\RR ds \int\hotimes\; \imath$.
}\end{ex}

\begin{ex}[Canonical $\;3$-trace for the Heisenberg group
  algebra] 
{\rm We first
construct the canonical $2$-trace of the group algebra of $\RR^2$ which
is equal to $\CC\rtimes_\aco\RR\rtimes_\aco\RR\cong SS\CC$.
On $\CC$ we consider the trace $\Tr$, a $0$-cocycle which is the
character of the $0$-cycle $(\CC,0,\id)$. Then we apply the
construction of Example~\ref{prop3} twice to obtain the double
suspension of $\Tr$ for the algebra
$SS\CC\cong \CC\rtimes_\aco\RR\rtimes_\aco\RR$, namely
$(\CC^{ss},d^{ss},\Tr^{ss})$
where $\CC^{ss}=SS\CC\hotimes\Lambda\Complex^2$, $d^{ss}=
\partial_{a_3}\hotimes e_1+\partial_{a_2}\hotimes e_2$ and
$\Tr^{ss}= \int_{\RR^2} da_2da_3 \Tr\otimes\imath$ 
(we use the notation of Example~\ref{ex5a}). 
If restricted to the smooth group algebra of $\RR^2$ this $2$-trace 
becomes a genuine cocycle referred to as 
the canonical cocycle of the group algebra of $\RR^2$. 
Its character is the well-known Chern character.

\vspace{.2cm}

Recall that the Heisenberg group algebra is isomorphic to 
$SS\CC\rtimes_\ttau\RR$
and so we seek to apply Theorem~\ref{prop2} to 
the action $\tilde\tau$ on the above double suspension. 
Since this action does not commute with $\partial_{a_3}$,
we cannot extend it trivially to the Grassmann generators. Instead we set
\begin{eqnarray*}
(\ttau_{a_1} (f\hotimes 1))(a_2)(a_3) &=&
f(a_2)(a_3-a_2a_1)\hotimes 1\;, \\
\ttau_{a_1}(1\hotimes e_1) &=& 1\hotimes  e_1\;, \\
\ttau_{a_1}(1\hotimes e_2) &=& 1\hotimes e_2 - a_1(1\hotimes e_1)\;.
\end{eqnarray*}
We claim that $\ttau_{a_1}$ commutes with $d^{ss}$: it suffices to
check this for elements of degree $0$ in $e_1$ and $e_2$ where we get
\begin{eqnarray*}
\ttau_{a_1}d^{ss}(f\hotimes 1)({a_2})(a_3) &= &
\partial_1 f({a_2})(a_3-{a_2} {a_1})\hotimes e_1 +
\partial_2 f({a_2})(a_3-{a_2} {a_1})\hotimes (e_2 -{a_1} e_1)\;,\\
d^{ss} \ttau_{a_1}(f\hotimes 1)({a_2})(a_3) &= &
\left(\partial_1 f({a_2})(a_3-{a_2} {a_1})-{a_1}\partial_2
  f({a_2})(a_3-{a_2} {a_1}) \right)\hotimes\, e_1 \\
&&+\: \partial_2 f({a_2})(a_3-{a_2} {a_1})\hotimes e_2 \;.
\end{eqnarray*} 
Furthermore, $\ttau$ leaves the graded trace $\Tr^{ss}$ invariant,
because of $\ttau_{a_1}(1\hotimes e_1 e_2)=1\hotimes e_1 e_2$ and the
translation invariance of the Lebesgue measure on $\RR^2$.
In order to apply Theorem~\ref{prop2}, we show that
the $2$-cycle $(\CC^{ss},d^{ss},\Tr^{ss})$ satisfies the
uniform bound (\ref{eq-ntracebound}) w.r.t.\ the dense
sub-algebra $C^1_c(\RR^2,\CC)\subset SS\CC$ given by continuously
differentiable functions with compact support. One finds for the norms
$p(f_1,f_2)$ ({\sl cf.} Definition~\ref{def3}), using
$\partial_{a_2}\ttau_t(f_1)=\ttau_t(\partial_1
f_1)-t\,\ttau_t(\partial_2 f_1)$
and $\partial_{a_3}\ttau_t(f_1)=\ttau_t(\partial_2 f_1)$, 
\begin{eqnarray*}
p(\ttau_t(f_1),f_2)
& \leq & 
\left(\|\ttau_t(\partial_1 f_1)\|\,\|\partial_2 f_2\|
\;+\; \|\ttau_t(\partial_2 f_1)\|\,\|\partial_1
f_2\|\right)|\supp(f_2)|
\\
& & +\; \|\ttau_t(\partial_2 f_1)\|\,\|\partial_2
f_2\|\;|t|\;|\supp(\ttau_tf_1)\cap\supp(f_2)|
\;. 
\end{eqnarray*}
Here $\|.\|$ is the supremum norm on $SS\CC$. Since
$|t|\:|\supp(\ttau_tf_1)\cap\supp(f_2)|$ 
is bounded in $t$ for any two compactly supported functions $f_1,f_2$, 
we obtain the
desired result, namely that $p(\ttau_t(f_1),f_2)$ is bounded in $t$.
Hence we are in a position to apply Theorem~\ref{prop2} 
from which we then infer a $3$-cycle over
$L^1(\RR,SS\CC,{\ttau})$ given by 
$$
\left(
L^1(\RR,SS\CC\hotimes\Lambda\CC^2,{\ttau})
\hotimes\Lambda\Complex, d^{ss}_{\ttau}, 
 \int_{\RR^2} d{a_2} da_3\, \Tr\, \eva_0 \otimes\imath
\right)\;.
$$
Its character is the canonical $3$-trace of the
$L^1$-crossed product $L^1(\RR,SS\CC,\ttau)$. 
The latter is dense in the Heisenberg group algebra and closed under
holomorphic functional calculus.
}\end{ex}

\begin{ex}\label{ex11}
{\rm If $(\Omega,d,\int)$ is an $\alpha$-invariant $n$-cycle over 
$\Bb$, then the above construction straightforwardly generalizes to
the $C^*$-field  $SS\Bb\rtimes_{\ttau\otimes\alpha}\RR$ from
Example~\ref{ex5} and we obtain the $n+3$-cycle
$$
\left(L^1(\RR,SS\Omega\hotimes\Lambda\CC^2,{\ttau\otimes\alpha})
\hotimes\Lambda\Complex, d^{ss}_{\ttau\otimes\alpha}, 
 \int_{\RR^2} da_2da_3 \int \eva_0 \otimes\imath
\right)\;.
$$
whose character is an $n+3$-trace on the Banach algebra
$L^1(\RR,SS\Bb,\ttau\otimes\alpha)$. The details of the proof that the
intermediate $n+2$-cycle is $\ttau\otimes\alpha$-invariant are
worked out as above and we only indicate how to show the
bound (\ref{eq-ntracebound}) for the $n+2$-cycle 
$$
\left(SS\Omega\hotimes\Lambda\CC^2, {d^{ss}}, 
 \int_{\RR^2} da_2da_3 \int \otimes\,\imath
\right)\;.
$$
For that consider the dense sub-algebra 
${\mathcal V}^{\fin}=\bigcup_{V\cup\Bb',\,\dim V<\infty}C_c^1(\RR^2,V)$ 
of $SS\Bb$. Then 
$$
\tilde Q(f_1,\dots,f_n):=|\supp(f_1)|\dots
|\supp(f_n)|\sup_{s_j,t_j\in\RR^2}
p(\alpha_{s_1}(f_1(t_1)),\dots,\alpha_{s_n}(f_n(t_n)))
$$  
is finite for all $f_j\in {\mathcal V}^{\fin}$,
and we get the desired bound (\ref{eq-ntracebound}) from
\begin{eqnarray*}
p(\ttau\otimes\alpha_t(f_1),f_2,\dots,f_{n+2})&\leq& 
\sum_{k\neq j}\|\partial_1 f_k\|\,\|\partial_2 f_j\|
\tilde Q(\dots\mbox{\rm no}\, f_k\,\mbox{\rm \&}\,f_j\dots)\\
&+& 
\sum_{j\neq 1}\|\partial_2 f_1\|\,\|\partial_2 f_j\|
\tilde Q(\dots\mbox{\rm no}\, f_1\,\mbox{\rm \&}\, f_j\dots)
|t||\supp(\ttau_tf_1)\cap\supp(f_j)|\,.
\end{eqnarray*}
}\end{ex}

\subsection{Chains and boundaries}

The interest in the following definition is that the graded traces are
not supposed to be closed.

\begin{defn}
An $n$-dimensional chain $(\Omega,d,\int,\partial\Omega,r)$ is a
graded differential algebra $(\Omega,d)$
of top degree $n$
together with a graded trace $\int$ on $\Omega^n$
and a surjective homomorphism of graded algebras of degree zero 
$r:\Omega\to \partial \Omega$ onto a graded algebra 
of top degree $n-1$ such that $d\ker( r)\subset \ker (r)$ and
$\int d\omega =0$ if $r(\omega)=0$. The chain is called a chain over $\Bb$ if
there exists an algebra homomorphism $\Bb\to \Omega^0$.

For a chain  $(\Omega,d,\int,\partial\Omega,r)$ over a Banach algebra
we require $r:\Omega\to \partial \Omega$ to be a continuous map
between Banach algebras. Such a chain is called unbounded if
$d$ and $\int$ and the character of the chain,
 $(A_0,\cdots,A_n)\mapsto \int  A_0dA_1 \cdots
dA_n$, are densely defined but satisfy condition (\ref{topcond}).
\end{defn}

As for cycles,
 we consider here only the case that $\Bb\subset \Omega^0$.

\vspace{.2cm}

An example of an unbounded chain for a commutative algebra $C(M)$ of
functions over a compact manifold is obtained when one looks at the
algebra of exterior forms with its usual differential and integration
structure, 
but $M$ has a boundary $\partial M$. 
The map $r$ is then simply the
restriction to the boundary and $\partial\Omega=\Omega(\partial M)$.
In this context, Stokes' Theorem relates integration of exact $\dim(M)$-forms
over $M$ to the integration of a form
over the boundary $\partial M$.
The following definition is motivated so 
that such a theorem holds automatically in the non-commutative setting.

\begin{defn} The boundary of an
$n$-dimensional chain  $(\Omega,d,\int, \partial \Omega,r)$ is the
$n-1$-dimensional cycle $(\partial \Omega,d',\int')$ where 
$$d'\omega'\;=\;rd\omega\;,
\qquad \int'\omega'\; =\; \int d\omega\;,
$$
for some $\omega\in r^{-1}(\omega')$.
\end{defn}

The following example is important for the construction of the dual 
of the ENN-map in
(\ref{ENN-map}).

\begin{ex}\label{ex-chain} {\rm
Consider the continuous $C^*$-field 
$SS\Bb\rtimes_{\ttau\otimes\alpha}\RR$ from Example~\ref{ex5} together
with an $\alpha$ invariant $n$-cycle $(\Omega,d,\int)$ over $\Bb$. 
In Example~\ref{ex11} we have constructed an $n+3$-cycle for a dense
Banach sub-algebra of the field. 
Now restrict this
$C^*$-field to the interval $[\hbar_0,\hbar_1]\subset \RR$, i.e.\ consider
$C_0([\hbar_0,\hbar_1]\times\RR,\Bb)\rtimes_{\ttau\otimes\alpha}\RR$. If we
repeat the construction of the $n+3$-cycle, we end up with a
graded differential algebra with graded trace
$$
\left(L^1(\RR,
C_0([\hbar_0,\hbar_1]\times\RR,\Omega)\hotimes\Lambda\CC^2,
{\ttau\otimes\alpha})\hotimes\Lambda\Complex\,, 
d^{ss}_{\ttau\otimes\alpha}, 
 \int_{[\hbar_0,\hbar_1]\times\RR} d\hbar \,ds \int \eva_0 \otimes\imath
\right)
$$
which is not closed. This is a chain if we set
$$
\partial \left(L^1(\RR,
C_0([\hbar_0,\hbar_1]\times\RR,\Omega)\hotimes\Lambda\CC^2,
{\ttau\otimes\alpha})\hotimes\Lambda\Complex\right) 
\;=\; 
\bigoplus_{j=0,1}L^1(\RR,C_0(\RR,\Omega)\hotimes\Lambda\CC,
{\tau^{\hbar_j}\otimes\alpha})\hotimes\Lambda\CC ,$$
the first Grassmann part
$\Lambda\Complex$ being the sub-algebra of $\Lambda\Complex^2$ of
elements not containing $e_1$,
and 
$$r(\omega) = \left\{\begin{array}{ll}
\omega|_{\hbar=\hbar_0}\oplus\omega|_{\hbar=\hbar_1}&
 \mbox{if } \omega e_1 \neq 0 \\ 
& \\
0 &\mbox{otherwise.}
\end{array}\right. $$
To verify this let us split   
$d^{ss}_{\ttau\otimes\alpha}=\delta_1+\delta_2$ where
$\delta_1(F\hotimes 1) = \partial_\hbar F \hotimes e_1$. Then
\begin{equation} \label{boundary}
 d^{ss}_{\ttau\otimes\alpha} F_1 \cdot\cdot\cdot 
d^{ss}_{\ttau\otimes\alpha} F_{n+3}\; =\;
\delta_1 (F_1 d^{ss}_{\ttau\otimes\alpha} F_2 \cdot\cdot\cdot d^{ss}_{\ttau\otimes\alpha}
F_{n+3}) +
\delta_2( F_1 d^{ss}_{\ttau\otimes\alpha} F_2 \cdot\cdot\cdot d^{ss}_{\ttau\otimes\alpha}
F_{n+3})\;.
\end{equation}
The graded trace $\int_{[\hbar_0,\hbar_1]\times\RR}\int \eva_0$ 
applied to the second term vanishes, because
$\int_{\RR}\int \eva_0$ is closed w.r.t.\ $\delta_2$.
The graded trace applied to the first term yields, when first
integrated over $[\hbar_0,\hbar_1]$ the boundary term, 
\begin{equation} \nonumber
\label{stokes}
\int_{[\hbar_0,\hbar_1]\times\RR}\int \eva_0 
\left(d^{ss}_{\ttau\otimes\alpha} F_1  \cdot\cdot\cdot
d^{ss}_{\ttau\otimes\alpha} F_{n+3}\right) 
\;=\; 
\left. \int_{\RR}\int  \eva_0
\left(F_1 \delta_2 F_2  \cdot\cdot\cdot
\delta_2 F_{n+3}\right) \right|^{\hbar_1}_{\hbar_0}.
\end{equation}
The integrand vanishes if $F_1 d^{ss}_{\ttau\otimes\alpha} F_2
\cdots d^{ss}_{\ttau\otimes\alpha} 
F_{n+3} $ lies in the kernel of $r$. 
This proves that
the above is indeed a chain, namely it shows that Stokes' Theorem
holds. Since this chain is essentially the restriction of the cycle 
of Example~\ref{ex11} to a closed interval, its character satisfies 
(\ref{topcond}) w.r.t.\ to the Banach sub-algebra  
$L^1(\RR,C_0([\hbar_0,\hbar_1]\times\RR,\Bb),{\ttau\otimes\alpha})$.
}\end{ex}

\section{Pairings between $K$-theory and higher traces}
\label{sec-pairing}

A cyclic $n$-cocycle $\eta$ over $\Bb$ extends to one
over $\Bb^+=\Bb\times\CC$ via
$\eta((A_0,\lambda_0),\dots,(A_{n},\lambda_n))=\eta(A_0,\dots,A_{n})$. 
Moreover, 
$\Tr\otimes\eta$ is a cyclic cocycle over $M_m(\Bb)\cong
M_m(\Complex)\otimes\Bb$ where $\Tr$ is the standard matrix trace. 
Define, for
a projection $p\in M_m(\Bb^+)$ or a $u\in U(M_m(\Bb))$, respectively,
\begin{eqnarray}\label{even}
\pair{\eta}{p}&=&c_{n}\; \Tr\otimes \eta(p,\ldots,p)
\mbox{ , }
\qquad
\mbox{$n$ even}\\ \label{odd}
\pair{\eta}{u}&=&
c_{n}\; \Tr\otimes \eta(u^*-1,u-1,\ldots)\mbox{ , }
\qquad
\mbox{$n$ odd}\;, 
\end{eqnarray}
the last formula with alternating entries. The
normalization constants are (chosen as in 
\cite{Pimsner83}, but not as in \cite{ConnesBook}):
$$
c_{2k}
\;=\;
\frac{1}{(2\pi\imath)^k}\,\frac{1}{k!}
\mbox{ , }
\qquad
c_{2k+1}
\;=\;
\frac{1}{(2\pi\imath)^{k+1}}\,
\frac{1}{2^{2k+1}}\,
\frac{1}{(k+\frac{1}{2})(k-\frac{1}{2})
\cdots\frac{1}{2}}
\mbox{ . }
$$

\subsection{General properties of pairings}
\label{sec-pairinggeneral}
The map $\pair{.}{.}$ defined in (\ref{even}) and (\ref{odd}) is refered to as
Connes' pairing between the $K$-theory and cyclic
cohomology of $\Bb$ because of the following property:

\begin{thm} {\bf{\rm \cite{Connes86,ConnesBook}}} 
Let $\Bb$ be a Banach algebra. The map 
$\pair{.}{.}$ induces bi-additive maps 
$K_n(\Bb)\times \bigoplus_{m\geq 0} HC^{2m+n}(\Bb) \to \CC$.
In particular, if
$\eta$ a cyclic $n$-cocycle and $x$
a projection in $M_m(\Bb^+)$ if $n$ is even, or a unitary of
$U(M_m(\Bb))$ if $n$ is odd, then
$\langle \eta,x\rangle$ depends only on the homotopy class
of $x$.
\end{thm}

Another point of view of this theorem is that
an even (odd) cyclic cocycle defines a functional 
$K_n(\Bb_0)\to \CC$, $n=0$ ($n=1$).
In Section~\ref{sec-cycles} we mentioned that
there are not enough (bounded) cyclic cocycles on a \CA\ and therefore
discussed $n$-traces on Banach algebras. These do equally well in this
context as the following proposition shows.

\begin{prop} {\bf{\rm \cite{Connes86}}}
If $\Bb$ is a
Banach algebra which is densely included in the \CA\ $\Cc$, then any
$n$-trace on $\Bb$ defines by extension of the formulas 
{\rm (\ref{even})} and {\rm (\ref{odd})} a functional on $K_n(\Cc)$.
\end{prop} 

The reason for this is that the algebra $\Bb''$ mentionned in
Section~\ref{sec-cycles} to which the $n$-trace can be extended by
continuity is closed under holomorphic functional calculus. 
This implies that $K_n(\Bb'')\cong K_n(\Bb)\cong K_n(\Cc)$ 
with isomorphisms induced by $\Bb''\subset\Bb\subset \Cc$ and
therefore any
class in $K_n(\Cc)$ contains a representative descending from $\Bb''$ 
which can be used to determine the
pairing.
In particular, the $n+1$-trace $\#_\alpha\eta$
constructed in Theorem~\ref{prop2}
yields a well-defined functional on the $K$-groups of the $C^*$-crossed
products. 

\vspace{.2cm}


Furthermore, Connes analysis of
\cite{Connes86} in which he shows that the character of 
a higher trace on $\Bb$ 
can be extended by continuity to a dense sub-algebra 
$\Bb''$ which is closed under holomorphic functional calculus does not
require $\int$ to be closed and therefore extends to the case of
chains. This implies that
the character of the boundary of an unbounded chain
$(\Omega,d,\int,\partial\Omega,r)$
over $\Bb=\Omega^0$ is fully defined on $r(\Bb'')$.
By continuity of the surjection
$r$, $r(\Bb'')$ is dense in $\partial\Omega^0$ and 
closed under holomorphic functional calculus. Hence 
$K_n(\Bb'')\cong K_n(\partial\Omega^0)$ with isomorphism induced by
inclusion. Therefore, the character $\eta$ of the boundary of the
unbounded chain defines by extension of the formulas 
{\rm (\ref{even})} and {\rm (\ref{odd})} a functional on 
$K_n(\partial\Omega^0)$.

\begin{prop}
\label{prop-chain}
Let $(\Omega,d,\int, \partial \Omega,r)$ be an $n$-dimensional {\rm (}possibly
unbounded{\rm )} 
chain over a Banach algebra $\Omega^0$ 
and consider $x'\in \partial\Omega^0$,
a projection if $n-1$ is even (a unitary if $n-1$ is odd). 
If there exists a projection $x\in\Omega^0$ if $n-1$ is even (a unitary
if $n-1$ is odd) such that $x'$ is homotopic to $r(x)$ then
$x'$ pairs trivially with the character $\eta$
of the boundary of the chain.
\end{prop}
\bew\ Consider first the case in which 
$n-1=2k$ and $x'$ is homotopic to $r(p)$ for a projection
$p\in\Omega^0$. Since the chain is unbounded $p$ can be found in a
sub-algebra of $\Omega^0$ which is closed under holomorphic functional
calculus and to which the character of the chain extends by
continuity. Then the pairing reads
$$\frac{1}{c_{2k}}\pair{\eta}{x'} \;=\; \int'r(p)(d'r(p))^{2k} 
\;=\; \int(dp)^{2k+1}
\;.
$$
As $\int$ is a graded trace, this
vanishes because
$$(dp)^{2k+1}\;=\;
p(dp)^{2k+1}(1-p)+(1-p)(dp)^{2k+1}p
\mbox{ . }
$$ 
If the degree is $n-1=2k+1$ we let $u$ be a unitary in the above dense
sub-algebra such that $x'$ is homotopic to $r(u)$. Then 
$$
\frac{1}{c_{2k+1}}\pair{\eta}{x'}\; =\; 
\int(du^*du)^{k+1} 
\;=\; - \int(dudu^*)^{k+1}\,.$$
The last equality follows because
$\int$ is a graded trace and the degree of $du$ and
$(du^*du)^{k}du^*$ are both odd. On the other hand, recall
$du^*=-u^*duu^*$ so that, using cyclicity again 

$$ \int(du^*du)^{k+1}\; =\; \int(-u^*duu^*du)^{k+1} \;=\; 
\int(-duu^*du u^*)^{k+1}\; = \;\int(dudu^*)^{k+1}
\mbox{ , }
$$

\noindent which shows that $\pair{\eta}{x'}$ has to vanish.\eb

\vspace{.2cm}

Under pairing with $K$-theory, cyclic
cohomology behaves like a periodic cohomology theory. 
This means, in
particular, that there exists a map 
(denoted $S$ in \cite{ConnesBook}) that assigns to
each cyclic $n$-cocycle a cyclic $n+2$-cocycle which pairs in the same
way with $K$-theory. In our context this reads as follows.

\vspace{.2cm}

Recall that an $n$-trace $\eta$ on $\Bb$ extends to an $n$-trace on the matrix
algebras over $\Bb$ or to $\K\otimes\Bb$ by the 
operator trace on the left factor, $\Tr\otimes\eta$. Furthermore,
following  \cite{Rieffel82} we construct in the appendix an isomorphism
$S\Bb\rtimes_{\tau\otimes\alpha}\RR\cong \K\otimes\Bb$ for any
$\RR$-action $\alpha$ on $\Bb$. If $\eta$ is $\alpha$-invariant then 
under this isomorphism $\Tr\otimes\eta$ gets identified
with the character $\eta^e$ of the $n$-cycle 
$(S\Omega\rtimes_{\tau\otimes\alpha}\RR,d^e,\int^e)$,
$$
(d^e f)(x)(s)\;=\;d(f(x)(s))
\mbox{ , }
\qquad
\int^e\omega\;=\;\int ds\int \omega(0)(s)
\mbox{ . }
$$ 
Although the following result is
known, we provide a proof of it in the appendix, filling in some details left
to reader in \cite{ConnesBook}.

\begin{thm}
\label{gleich2} 
Let $\eta$ be an $\alpha$-invariant $n$-trace on $\Bb$ and
$x$ be a representative for an element in 
$K_i(S\Bb \rtimes_{\tau\otimes\alpha}\RR)$. Then
\begin{equation}\label{eq-doubleperiod} 
\pair{\#_{\tau\otimes\alpha}\eta^s}{x}
\;=\; 
-\,\frac{1}{2\pi}\;\pair{\eta^e}{x} 
\mbox{ . }
\end{equation}
\end{thm}

\subsection{The duality equation}
\label{sec-main+coro}

In the following proposition we construct the dual to the ENN-maps
(\ref{ENN-map}). 
\begin{prop}\label{prop1c} 
Let $\eta$ be an $\alpha$-invariant $n$-trace on $\Bb$
and $SS\Bb\rtimes_{\ttau\otimes\alpha}\RR$
the \CF\ from {\rm Example~\ref{ex5}} with fibre map $\varphi_\hbar$. 
Furthermore, $x\in M_m((SS\Bb\rtimes_{\ttau\otimes\alpha}\RR)^+)$ is 
a projection if $n$ is even, or
$x\in U(M_m(SS\Bb\rtimes_{\ttau\otimes\alpha}\RR)$
if $\eta$ is odd. 
Denote $x_{\hbar}=\varphi_\hbar(x)$.
Then the pairings
$$
\pair{\#_{\tau^{\hbar}\otimes\alpha}\eta^s}{x_{\hbar}}
$$
{\rm (}for a given $\hbar$ over the algebra 
$S\Bb\rtimes_{\tau^\hbar\otimes\alpha}\RR${\rm )}
are independent of $\hbar$. In other words, the map
$$\#_{\tau^{1}\otimes\alpha}\eta^s\mapsto
\#_{\tau^{0}\otimes\alpha}\eta^s$$ 
is the dual to the {\rm ENN}-map
{\rm (\ref{ENN-map})}. 
\end{prop} 

\noindent
\bew\ We apply Proposition~\ref{prop-chain} to the chain constructed
in Example~\ref{ex-chain}. 
The boundary of this chain is a cycle for 
$S\Bb\rtimes_{\tau^{\hbar_0}\otimes\alpha}\RR\oplus 
S\Bb\rtimes_{\tau^{\hbar_1}\otimes\alpha}\RR$ and its character is
given by $\#_{\tau^{\hbar_1}\otimes\alpha}\eta^s\oplus 
-\#_{\tau^{\hbar_0}\otimes\alpha}\eta^s$, {\sl cf.} equation 
(\ref{boundary}). 
Furthermore 
$r(x)=(x_{\hbar_0},x_{\hbar_1})$ is a projection (unitary)
in that algebra. 
By Proposition~\ref{prop-chain} 
$(x_{\hbar_0},x_{\hbar_1})$  therefore pairs trivially
with the character of the boundary of the chain, i.e.
$0=\pair{
\#_{\tau^{\hbar_1}\otimes\alpha}\eta^s}{x_{\hbar_1}}-
\pair{\#_{\tau^{\hbar_0}\otimes\alpha}\eta^s}{x_{\hbar_0}}\;$.
\eb


\begin{thm}
\label{gleich} 
Let $x$ be a representative for an element in $K_i(\Bb\rtimes_{\alpha}\RR)$ 
and $\eta$ be an $\alpha$-invariant
$n$-trace over $\Bb$.
Then
\begin{equation}\label{equ1} \pair{\#_\alpha\eta}{x} 
\;=\; 
-\pair{\#_{\tau\otimes\alpha}\eta^s}{\partial_i(x)}
\mbox{ , }
\end{equation}
where $\partial_0=\mbox{\rm exp}:K_0(\Bb\rtimes_{\alpha}\RR)\to
K_1(S\Bb \rtimes_{\tau\otimes\alpha}\RR)$ and 
$\partial_1=\mbox{\rm ind}:K_1(\Bb\rtimes_{\alpha}\RR)\to 
K_0(S\Bb \rtimes_{\tau\otimes\alpha}\RR)$.
\end{thm}

\noindent\bew\ 
Applying Proposition~\ref{prop1c} to the points $\hbar=0$ and
$\hbar=1$ and Proposition \ref{prop-ENNWH}, it suffices to show that
$\pair{\#_\alpha\eta}{x} = - \pair{\#_{\idl\otimes\alpha}
\,\eta^s}{\partial_i(x)}$ where now  $\partial_0=\beta$ and 
$\partial_1=\Theta$.

\vspace{.2cm}

Let us look here at the case $i=0$ in which $x$ is represented by a
projection $p$. Then $\beta[p]$ is represented by the unitary $Gp+1$ where 
$G(s)=e^{2\pi \imath \chi(s)} -1$. 
Since $G$ commutes with $p$,  the calculation is simple.
If $\eta$ is the character of the $2k+1$ cycle $(\Omega,d,\int)$, then
$\#_{\idl\otimes\alpha}\,\eta^s$ is the character of
$(\Omega^s_\alpha,d^s_\alpha,\int_\RR\int \eva_0)$ and we write again
$d_\alpha^s = \delta_1 + d_\alpha$ where $\delta_1(f\hotimes 1) =
\partial_s f\hotimes e_2$. Then $d_\alpha^s (Gp) = (\partial_s G)
p\hotimes e_2+
G d_\alpha p$ so that, using $p (d_\alpha p)^m p = p (d_\alpha p)^m$
for even $m$ and $p (d_\alpha p)^m p = 0$ for odd $m$, one gets
\begin{eqnarray*} 
\pair{\#_{\idl\otimes\alpha}\eta^s}{Gp+1} 
&=& 
c_{2k+3}\int_\RR \int 
\eva_0\left(
\bar{G}p \,
d_\alpha^s(Gp)\left(d_\alpha^s(\bar{G}p)d_\alpha^s(Gp)\right)^{k+1}
\right) 
\\ 
&=& -\;(k+2)c_{2k+3}\int_\RR G'G^{k+1}\bar{G}^{k+2} \int 
\eva_0\left(p (d_\alpha p)^{2k+2}\right) \\
&=& -\; \pair{\#_\alpha\eta}{p}.
\end{eqnarray*}
since  $\int_\RR G'G^{k+1}\bar{G}^{k+2}=2\pi \imath \;
\frac{(2k+3)!}{(k+1)!(k+2)!}\,$.

\vspace{.2cm}

As the case $i=1$ is not
needed for our application to the quantum Hall effect, 
we refer the reader to \cite{Pimsner83} for the corresponding
calculation. \eb


\begin{thm}
\label{theo-duality} Consider an
$\RR$-action $\alpha$ on a \CA\ $\Bb$ together with an
$\alpha$ invariant $n$-trace $\eta$ on $\Bb$. Then 
$\#_\alpha\eta$ satisfies the duality equation
$$
\pair{\#_\alpha\eta}{x}
\;=\;
-\,\frac{1}{2\pi}\;\pair{\eta}{\partial_i x}
\;,
\qquad x\in K_i(\Bb\rtimes_\alpha\RR)
\;,
$$
where $\partial_i$ are the boundary maps of
$K$-theory associated to {\rm(\ref{eq-introext})}. 
\end{thm}

\noindent
\bew\ 
Combine Theorem~\ref{gleich} and Theorem~\ref{gleich2}. \eb

\section{Topology of the integer quantum Hall effect}
\label{sec-equal}

In quantum Hall samples, there are two current carrying mechanisms:
edge currents flow along the boundaries due to intercepted cyclotron
orbits and bulk currents result from the Lorentz drift in presence of
an exterior electric field. Important in the present context is that
both these currents are topologically quantized by a Fredholm index
resulting from a pairing of adequate elements
of K-groups and higher traces (unbounded cyclic cocyles). 
While for bulk currents, this was known
for a long time \cite{Bel85,AS85,K87,Bellissard88,ASS,BES}, quantisation of edge
currents was only proven more recently and under a gap condition on
the bulk Hamiltonian, namely, in a tight binding context in
\cite{SKR,KRS02,ElbauGraf02}, and for continuous magnetic
(differential) operators in \cite{KS1}. 
In the continuous case,
the two pairings turn out to be over two of the algebras in the
Wiener-Hopf extension (\ref{eq-introext}) and Theorem
\ref{theo-duality} then implies equality of bulk and edge Hall
conductivities. To explain this result and the quantities involved 
in more detail is the subject of this section. 

\vspace{.2cm}

Within the tight-binding approximation 
an analogous result was proven
in \cite{KRS02} and re-derived by ad-hoc methods by Elbau and Graf
 \cite{ElbauGraf02}.
Equality of bulk and edge conductivities appeared already in various
other guises, for instance, in the framework of scaling theory \cite{Pru} and 
that of classical mechanics as resulting from a simple 
conservation law \cite{Fro}.

\subsection{Bulk and edge-Hall conductivity}  
\label{sec-Hallpairings}

Here we summarize the framework and the results
of \cite{KS1} and then state the main
theorem for quantum Hall systems.
We described a quantum Hall system with disorder and a boundary by
means of covariant families of integral operators on $L^2(\RR^2)$.
For this we combined a Borel probability space $(\hull,\PP)$ whose
elements describe disorder configurations\footnote{We use here 
the custumory notation 
$\hull$ for the space of disorder configurations. It should
not be confused with the notation for graded algebras used earlier.}
in $\RR^2$ with the space
$\hat{\RR}=\RR\cup \{+\infty\}$ (topologically a half open interval). 
Points $s\in \Rh$ describe the position of
the boundary of the half-space\footnote{It is easier in the
present context to work with the left half space instead of
the right half space we used in \cite{KS1}.} 
$\RR\times\RR^{\leq s}$ where $\RR^{\leq s}=\{x\in\RR|x\leq s\}$. 
The space $\hull$ carries an $\RR^2$-action which is reminiscent of 
the translation of a disorder configuration. We denote this action by
$\om\mapsto \x\cdot\om$. For concreteness we may think of $\x\cdot\om$
as the configuration $\om$ being shifted by $\x$, {\sl i.e.}
$\x\cdot\om$ looks at $\y+\x$ like $\om$ at $\y$.
The probability measure $\PP$ is required to be invariant and ergodic
under this action. Furthermore, $\Omega$ carries a compact metric
topology w.r.t.\ which the action is continuous.
The $\RR^2$-action is extended to an action on 
$\hat\hull=\hull\times(\RR\cup \{+\infty\})$ by 
$(\om,s)\mapsto (\x\cdot\om,s+x_2)$, that is the extended action shifts
the boundary in the same direction as the configuration.
A family $A=(A_{\hat{\om}})_{\hat{\om}\in\hat{\hull}}$
of integral operators on $L^2(\RR^2)$ is called covariant if

\begin{equation}
\label{eq-cov}\nonumber
U(\vec \xi\,)A_{\hat{\om}} U(\vec \xi\,)^* 
\;=\; A_{{\vec \xi}\cdot {\hat{\om}}}
\mbox{ , }
\qquad
\vec \xi\in\RR^2
\mbox{ , }
\end{equation}

\noindent where $U(\vec \xi\,)$ are the magnetic translation operators
defined in Section \ref{obsalg} below.
The quantum Hall system is described by a random family 
$H=(H_{\hat{\omega}})_{\hat{\om}\in\hat{\hull}}$ of Hamiltonians, 
$$H_{\hat{\omega}}=
\frac{\hbar^2}{2m}((\imath\partial_1 - \gamma
X_2)^2+\imath\partial_2^2) + V_\om
$$
acting on $L^2(\RR\times\RR^{\leq s})$
with Dirichlet boundary conditions at $s$.
Here $V_\om$ is the potential
depending on the random variable $\om\in\hull$
and $\gamma$ is the strength of the magnetic field.
It is shown in \cite{KS1} that
sufficiently regular bounded and compactly supported functions of $H$ yield
covariant families of integral operators. 
Moreover, if we push the boundary to ${+\infty}$,
then we describe a disordered system without boundary by a family of 
Hamiltonians
denoted $H_\infty=(H_{(\omega,\infty)})_{{\om}\in{\hull}}$. 
In this framework \cite{BES,ASS}, the bulk Hall conductivity of a gas of
independent electrons described by the planar Hamiltonian 
at zero-temperature
and with chemical potential $\mu$ belonging to a mobility gap 
in the spectrum of the Hamiltonian is given by
\begin{equation}\label{ch2}
\sigma^\perp_b(\mu)
\;=\;
\frac{q^2}{h}\;\chd(P_\mu,P_\mu,P_\mu)\;,\quad
\chd(A,B,C)= -2\pi\imath \;
\TV (A([X_1,B][X_2,C]-[X_2,B][X_1,C]))\;.
\end{equation}
Here $P_\mu=\chi_{(-\infty,\mu]}(H_\infty)$ 
is the covariant family of associated Fermi projections, $X_j$ the
$j$-component of the position operator and $\TV$ the trace per unit
volume (precise definition given in eq.~(\ref{eq-tau}) below).

\vspace{.2cm}

The main result of
\cite{KS1} is that 
the edge Hall conductivity of a gas of
independent electrons described by the Hamiltonian $H$ with Dirichlet
boundary conditions at the edge 
at zero-temperature and
with chemical potential $\mu$ belonging to a gap $\Delta$ in the
spectrum of the {\em planar} Hamiltonian $H_\infty$ is given by

\begin{equation}\label{wind}
\sigma^\perp_e(\mu)
\;=\;
-\,\frac{q^2}{h}\;\xi(\Uu^*(\Delta)-1,\Uu(\Delta)-1)\;,
\qquad 
\xi(A,B)= \TVh (A[X_1,B])\;.
\end{equation}
Here ${\cal U}(\Delta)$ is constructed from the half-planar Hamiltonian
via functional calculus
\begin{equation}
\label{def-unitary}
{\cal U}(\Delta)
\;=\;
\exp (-2\pi \imath \,G({H}))
\mbox{  }
\end{equation}

\noindent 
for a monotonously decreasing smooth function $G:\RR\to\RR$ with 
$G(-\infty)=1$, $G(\infty)=0$, and
$\mbox{\rm supp}(G')\subset \Delta\backslash G^{-1}(\frac{1}{2})$,
and $\TVh$ is the trace per unit length along the boundary combined
with the operator trace perpendicular to the boundary (its precise
definition is given in eq.~(\ref{eq-tauhat}) below).
The corollary of Theorem \ref{theo-duality} is then:

\begin{thm}
\label{theo-equality}
Suppose that $E$ is in a gap of $H_\infty$, the Hamiltonian on the
plane without boundary. Then
$$
\sigma^\perp_e(E) 
\;=\;\sigma^\perp_b(E) 
\mbox{ . }
$$
\end{thm}

In the Landau model and its restriction to the half-space,
a proof can be given by explicitly calculating both pairings and then
seeing that both numbers are the same \cite{SKR}. 
This calculation makes use of
the translation invariance in the direction along the boundary. 
Not only does it exclude any type of disorder
which breaks that invariance, it is also not very satisfactory in that
it does not show directly that both pairings are the same.  
Moreover, the explicit calculation of both sides becomes already very
complicated if a periodic potential is added.

\vspace{.2cm}

To prove the theorem we need to describe how to view covariant
families of operators as elements of \CA s and how 
(\ref{ch2}) and (\ref{wind}) can be interpreted as pairings.

\subsection{Observable algebra with disorder and boundary}
\label{obsalg}

The first component of the action $\om \mapsto \x\cdot \om$
of $\RR^2$ on $\hull$  corresponds to
translating the disorder configuration along the boundary and yields 
an $\RR$-action on
$C(\hull)$ given by
$$\aca_{x_1}(f)(\om) = f((x_1,0)\cdot\om)\;.$$
The second component of the action $\om\mapsto \x\cdot \om$
yields for given $\gamma\in\RR$ an $\RR$-action on the
crossed product $C(\hull)\rtimes_{\aca}\RR$
by ($f:\RR\to C(\hull)$)
$$
\acb_{x_2}(f)(x_1)(\om)\; =\; 
e^{\imath\gamma x_1x_2}f(x_1)((0,x_2)\cdot\om)\;.
$$
This corresponds to
translating the disorder configuration perpendicular to the boundary
together with a phase shift depending on $\gamma$. We can interprete
$\gamma$ as the strength of the magnetic field and
$\Aa=C(\hull)\rtimes_{\aca}\RR\rtimes_{\acb}\RR$
as the observable algebra for the planar model.

\vspace{.2cm}

Following the philosophy in \cite{KS1}, we can view $\Aa$ as
being obtained from a larger algebra
$\Ta=C_0(\hat\hull)\rtimes_{\aca}\RR\rtimes_{\tau\otimes\acb}\RR$
where we extend $\aca$ trivially to the second factor
of $\hat{\hull}$ and
$$(\tau\otimes\acb)_{x_2}(f)(x_1)(\hat\om) = 
\acb_{x_2}f(x_1)((0,x_2)\cdot\hat\om)\;.$$
$\Ta$ can be interpreted as the observable algebra for the system with
boundary. Its 
relation with the covariant operator families of \cite{KS1} is as
follows. 

\vspace{.2cm}

A point $\oh\in\hat\hull$ defines a one-dimensional representation of
$C_0(\hat\hull)$, $\rho_\oh: C_0(\hat\hull)\to\CC$:
$\rho_\oh(f)=f(\oh)$.
Applying (\ref{rep}) twice,
we get a representation $\pi_\oh$ of $\Ta$ on
$L^2(\RR^2)$. If $F:\RR\to(\RR\to C_0(\hat\hull))$, then the integral kernel
of $\pi_\oh(F)$ is
\begin{eqnarray*} 
\langle \x\,|\pi_\oh(F)|\y\,\rangle &=&\rho_\oh \aca_{-x_1} 
\left( (\tau\otimes\acb)_{-x_2} \left(F(x_2-y_2)\right)(x_1-y_1)\right)\\
&=& e^{-\imath\gamma(x_1-y_1)x_2} F(x_2-y_2)(x_1-y_1)(-\x\cdot \oh)\; .
\end{eqnarray*}
It follows that $\langle \x-{\vec \xi}\,|\pi_\oh(F)|\y-{\vec \xi}\,\rangle
= e^{\imath\gamma(x_1-y_1)\xi_2}\langle \x\,|\pi_{{\vec
    \xi}\cdot\oh}(F)|\y\,\rangle$ and hence
$$  U({\vec \xi}\,)\pi_\oh(F)U({\vec \xi}\,)^* \;= \;
\pi_{{\vec
    \xi}\cdot\oh}(F) \;,$$
where the magnetic translation operators $U({\vec \xi}\,)$ 
are defined by 
\begin{eqnarray*}
(U({\vec \xi}\,)\psi)(\x\,)
& = &
\hat{\Phi}({\vec \xi},\x-{\vec \xi}\,)\,\psi(\x-{\vec \xi}\,)
\mbox{ , }
\qquad
\hat{\Phi}({\vec \xi},\x\,)\;=\;e^{-\imath\gamma\xi_2x_1}
\mbox{ . }
\end{eqnarray*}
The collection $\pi(F)=(\pi_\oh(F))_{\oh\in\hat\hull}$ forms therefore 
a covariant family of bounded integral operators. 
By construction the operators are weakly
continuous in $\oh$ and the norm
$\|\pi(F)\|_\infty$ which we defined in \cite{KS1}
to be the essential supremum over $\|\pi_\oh(F)\|$ is bounded by the
$C^*$-norm $\|F\|$. If we apply the above construction to each summand in
the direct sum representation 
$\rho=\bigoplus_{\oh\in\hat\hull}\rho_\oh$ we obtain a representation
$\pi$ which also decomposes into a direct sum representation, namely
on $\bigoplus_{\oh\in\hat\hull}L^2(\RR^2)$, and we can interprete the
covariant family $(\pi_\oh(F))_{\oh\in\hat\hull}$ as the
representative $\pi(F)$ of $F$. Since the direct sum representation 
$\rho$ is faithful, also $\pi$ is faithful and so, first, identifies
$\Ta$ with a sub-algebra of the completion of the algebra
of weakly continuous covariant families of bounded integral operators
denoted
$\Ta$ in \cite{KS1}, and second, implies $\|\pi(F)\|_\infty=\|F\|$. 
The estimates established in \cite{KS1} now show that
for potentials $V_\om(\x):=V(-\x\cdot \om)$ with $V\in C(\hull)$
and differentiable along the flow of the $\RR^2$ action 
and $F\in C^k_c(\RR)$ with $k>6$, 
the covariant families $F(H)$ and $D_jF(H)$ can be viewed as
elements of $L^1(\RR,L^1(\RR,C_0(\hat\hull),\aca),\tau\otimes\acb)$ and
hence of $\Ta$.

\subsection{Pushing the boundary to infinity}\label{ex4} 
The second component in $(\om,s)\in \hat\hull$ describes the
  position of the boundary and is allowed to take the value ${+\infty}$.
The evaluation 
$\eva_\infty(F)(x_2)(x_1)(\om) =  F(x_2)(x_1)(\om,{+\infty})$ 
has the effect of pushing the boundary to ${+\infty}$. 
The algebra ${\Ta}_\infty =
C(\hull)\rtimes_{\aca}\RR\rtimes_{\acb}\RR$
is therefore the observable algebra of the model with disorder, but
without a boundary (planar model). 
Now the crucial observation is that 
the pushing of the boundary to infinity defines a
surjective algebra morphism
$$\eva_\infty:\Ta\to\Aa\mbox{ . }$$
Therefore $\Ta$ is an extension of $\Aa$ by the edge algebra
$\Ea:=\ker (\eva_\infty)$ which can be understood as the algebra of
observables which are located at the boundary. 
This gives an exact sequence precisely of the form
(\ref{eq-introext}).

\subsection{Chern class and non-commutative winding number}
\label{exch}

The Chern class $\ch$ of (\ref{ch2})
and the $1$-trace $\xi$ of (\ref{wind})
can be obtained by application of Proposition~\ref{prop1a}. 
In the first case the algebra is
${\Ta}_\infty = C(\hull)\rtimes_{\aca}\RR\rtimes_{\acb}\RR$,
with trace $\TV:\Aa\to\CC$:

\begin{equation} 
\label{eq-tau}
\TV(F)
\;=\;\int_\hull d\PP(\om) \;F(0)(0)(\om)
\end{equation}
and derivations 
\begin{equation}\nonumber 
\nabla_j(F)(x_2)(x_1)
\;=\;
\imath x_j F(x_2)(x_1)\,.
\end{equation} 
Then $\chd$ is then $-2\pi \imath$ times the character of the $2$-cycle
constructed from these data using Proposition~\ref{prop1a}.
Its domain includes the dense sub-algebra $C_c(\RR,C_c(\RR,C(\hull)))$.

\vspace{.2cm}

In the second case, the algebra is the ideal
$\Ea=\ker(\eva_\infty)\subset\Ta=
C_0(\hat\hull)\rtimes_{\aca}\RR\rtimes_{\tau\otimes\acb}\RR$ 
with trace $\TVh:\Ta\to\CC$:
\begin{equation} 
\label{eq-tauhat}
\TVh(F)
\;=\;
\int_{\hat\hull} d\PP(\om)ds\; F(0)(0)(\om,s)
\end{equation}
and derivation
\begin{equation} \nonumber
\nabla_1(F)(x_2)(x_1)
\;=\;
\imath x_1 F(x_2)(x_1).
\end{equation}
The character of the $1$-trace
constructed from these data using Proposition~\ref{prop1a} is $\xi$
and its domain contains $C_c(\RR,C_c(\RR,C(\hat\hull)))\cap \Ea$.

\subsection{Proof of Theorem~\ref{theo-equality}}
Theorem~\ref{theo-equality} is obtained by application of
Theorem~\ref{theo-duality}
to the algebras and actions involved in Section~\ref{obsalg} and the
higher traces from Section~\ref{exch}. Specifically, we take
$\Bb=C(\Omega)\rtimes_{\aca}\RR$ with action 
$\alpha=\acb$ and 1-trace $\eta$ on $\Bb$ given by
$\eta(f,g)= \int_\hull d\PP(\om)\, (f{\nabla}_{1}g)(0)(\om)$.
Then $\xi$ from (\ref{wind}) is given by $\xi=\imath \eta^e$ and $\chd$
from (\ref{ch2}) by $\chd=-2\pi\imath \;\#_{\alpha}\eta$. Now let 
$P_\mu\in \Aa=\Bb\rtimes_{\alpha}\RR$ be the element corresponding
to the Fermi projection. The exponential map associated with  
the extension defined by
$\Ta\stackrel{\eval_\infty}{\longrightarrow}\Aa$
yields $[\Uu(\Delta)]_1=\exp[P_\mu]_0$. Thus  
$$ 
\frac{1}{2\pi \imath}\; \langle \chd,P_\mu\rangle
\; =\; 
\langle
\#_{\tau\otimes\alpha}\eta^s,\Uu(\Delta)\rangle 
\;=\;
-\, \frac{1}{2\pi \imath}\;  \langle\xi,\Uu(\Delta)\rangle
\;,
$$
the first equality following from Theorem~\ref{gleich} and
the second from Theorem~\ref{gleich2}. Since
$c_1=c_2$ 
we have 
$$
\chd(P_\mu,P_\mu,P_\mu) \;=\; - \,\xi(\Uu(\Delta)^*-1,\Uu(\Delta)-1)
\;,
$$
which proves Theorem~\ref{theo-equality}.
\eb

\begin{appendix}

\section{Periodicity in cyclic cohomology}
\label{sec-suspension}

For the convenience of the reader we present a detailed proof of 
Theorem~\ref{gleich2}.
The following two isomorphisms \cite{Rieffel82} allow to
reduce this proof to a calculation for compact operators:

\begin{equation}
\label{Psi}
\Psi:S\Bb \rtimes_{  \tau\otimes\alpha}\RR \to  S\Bb\rtimes_{
  \tau\otimes\idl}\RR
\mbox{ , }
\qquad
\Psi(f)(x)(s)\; =\; \alpha_{s}(f(x)(s))
\mbox{ , }
\end{equation}

\noindent and (identifying $S\Bb \rtimes_{\tau\otimes\idl}\RR$ with
$S\Complex \rtimes_{  \tau}\RR\otimes\Bb$)

\begin{equation}
\label{Phi}
\Phi:S\Bb \rtimes_{  \tau\otimes\idl}\RR \to \K(L^2(\RR))\otimes \Bb
\mbox{ , }
\qquad
 \Phi\;=\;
\rho\otimes\id
\mbox{ , }
\end{equation}

\noindent where $\rho$ is the 
representation of $S\Complex \rtimes_{\tau}\RR$ on $L^2(\RR)$ given by
$$
(\rho(f)\psi)(x) \;=\; 
\int dy\, f(y)(x)\,\psi(x-y)
\;=\; 
\int dy \,f(x-y)(x)\,\psi(y)
\mbox{ . }
$$
Hence the integral kernel of $\rho(f)$ is $\langle x|\rho(f)|y\rangle
= f(x-y)(x)$ so that $\Tr(\rho(f))=\int_\RR dx f(0)(x)$.

\begin{lem}
\label{lem-pushs} 
Let $\eta$ be a $\alpha$-invariant cyclic cocycle over
$\Bb$. Then 
\begin{equation}\label{lemeq1}
\#_{\tau\otimes \alpha}\,\eta^s\;=\;\Psi^*\#_{\tau\otimes\idl}\,\eta^s
\mbox{ , }
\end{equation}
\begin{equation}\label{lemeq2}
\eta^e\;=\;\Psi^*\Phi^*\Tr\otimes\eta
\mbox{ . }
\end{equation}
\end{lem}

\noindent 
\bew\ 
Let $\eta$ be the character of $(\Omega,d,\int)$. 
One has
$$
\#_{\tau\otimes \alpha}\,\eta^s(f_0,\ldots,f_n) 
\;= \;
\int_\RR ds \int 
\left(f_0 d_{\tau\otimes\alpha}^s f_1\cdots d_{\tau\otimes\alpha}^s
f_n
\right)(0)(s)\mbox{ , }
$$
where the product is that in $S\Omega\rtimes_{\tau\otimes\alpha} \RR$.
On the other hand 
$$
(\Psi^*\#_{\tau\otimes \idl}\,\eta^s)(f_0,\ldots,f_n) 
= 
\int_\RR ds \int 
\left(\Psi(f_0)d_{\tau\otimes\idl}^s\Psi(f_1) \cdots
d_{\tau\otimes\idl}^s\Psi(f_n)\right)(0)(s),$$
with product in $S\Omega\rtimes_{{\tau\otimes\idl}} \RR$.
Now (\ref{lemeq1}) follows from $ d_{\tau\otimes\idl}^s\Psi(f)=
\Psi(d_{\tau\otimes\alpha}^s f)$ and
$$
\left(\Psi(f_0) d_{\tau\otimes\idl}^s\Psi(f_1) \cdots
d_{\tau\otimes\idl}^s\Psi(f_n)\right)(0)(s)
\;=\; \alpha_s 
\left(f_0 d_{\tau\otimes\alpha}^s f_1 
\cdots d_{\tau\otimes\alpha}^s f_n
\right)(0)(s)
$$ 
and the $\alpha$-invariance of $\eta$.

For (\ref{lemeq2}), let $f_j=\Psi^{-1}(g_j\otimes b_j)$ where 
$g_j\otimes b_j\in S\Complex\rtimes_{\tau}\RR\otimes \Bb\cong
S\Bb\rtimes_{\tau\otimes  \idl}\RR$.
Then 
\begin{eqnarray*}
\Psi^*\Phi^*\,\Tr\otimes\eta(f_0,\ldots,f_n) 
&= &\Tr\otimes\eta(\rho(g_0)\otimes b_0,\ldots,\rho(g_n)\otimes b_n)
\\
&=&
\Tr(\rho(g_0\cdots g_n)) \int b_0db_1\cdots db_n \;,\\
\eta^e(f_0,\ldots,f_n) 
&= &\int_\RR ds\int \eva_0 \:\Psi^{-1}(g_0\otimes b_0)
d\Psi^{-1}(g_1\otimes b_1)
\cdots d\Psi^{-1}(g_n\otimes b_n)\\ 
&=& \int_\RR ds\, \eva_0 (g_0\cdots g_n) \int
b_0 db_1\cdots b_n\;,
\end{eqnarray*}
and the lemma follows from the fact that $\Tr(\rho(g)) = \int_\RR ds\,
g(0)(s)$ for $g\in  S\Complex\rtimes_{\tau}\RR$.
\eb

\vspace{.2cm}

\begin{lem}\label{thm3}
Let $\eta$ be an $\alpha$-invariant cyclic cocycle over
$\Bb$. Then for a projection $X\in \K\otimes\Bb$ or a unitary $X\in
U(\K\otimes\Bb)$ one has
$$
\langle \Tr\otimes \eta, X\rangle  
\;=\; -\,2\pi\;
\langle\#_{\tau\otimes \idl}\eta^s,\Phi^{-1}(X)\rangle\;.
$$
\end{lem}

\noindent \bew\ 
We start with the case $\Bb=\Complex$ and $\eta=\Tr$ which
is the character of $(\CC,0,\Tr)$.
Then $\Tr\otimes \eta=\Tr$ and 
$\#_{\tau\otimes\idl}\eta^s=\#_{\tau}\Tr^s$ is the character of 
$(S\CC\rtimes_\tau\RR\otimes\Lambda\CC^2,\delta,\int^s_\tau)$
where, for  $f:\RR\to S\Complex$, $\delta=\delta_1+\delta_2$ with
$\delta_1(f) =\partial_s f\otimes e_1$,
$\delta_2(f) =\nabla_x f\otimes e_2$, $\nabla_xf(x)=\imath xf(x)$, and
$\int^s_\tau f\otimes e_1e_2 = \int_\RR ds f(0)(s)=\Tr(\rho(f))$.
Let $M$ and $D$ be operators given by 
$M\psi(x)=\imath x\psi(x)$ and $D\psi(x)=\psi'(x)$ with usual common
domain $C^1_c(\RR)$. They satisfy
$[D,M]=\imath$. 
Then, for differentiable $f$,
$$
\langle x|[M,\rho(f)]|y\rangle 
\;=\; \imath (x-y)f(x-y)(x) 
\;=\; 
\langle x|\rho(\nabla_x f)|y\rangle
\;,
$$ 
$$
\langle x|[D,\rho(f)]|y\rangle
\;=\; 
(\partial_x+\partial_y) f(x-y)(x)
\;=\; 
\langle x|\rho(\partial_s f)|y\rangle
\;,
$$
and therefore, if $p\in S\CC \rtimes_{  \tau}\RR$ is a differentiable 
projection,
\begin{equation}\label{eq10}
\langle\#_\tau\Tr^s,p\rangle 
\;= \;
\frac{1}{2\pi\imath}\; 
\Tr \left(\rho(p)[[D,\rho(p)],[M,\rho(p)]]\right) 
\;=\;  
-\,\frac{1}{2\pi}\;\Tr(\rho(p))\;.
\end{equation}
In the last equation, we used
$P[[D,P],[M,P]]P=-P[D,M]P + [PDP,PMP]$, $P=\rho(p)=P^2$. 
This proves the statement for $\Bb=\CC$ and $\eta=\Tr$.

\vspace{.2cm}

In the general case,
$\eta$ is the character of some $n$-cycle $(\Omega,d,\int)$ over
$\Bb$. Then
$d_{ \tau\otimes \idl}^s=d'+\delta$ where $\delta$ is as above 
and $d'(f)(x)(s)= d(f (x)(s))$. 

\vspace{.2cm}

We apply this first to a projection
of the form $X=\rho(p)\otimes x\in \K\otimes\Bb^+$ where 
$x$ and $p$ are projections. Then
$d_{ \tau\otimes \idl}^s(\Phi^{-1}(X)) = p\otimes d x +
\delta  p \otimes x$.
Let $n=2k$, $k\geq 0$. Using  $p(\delta
p)p^j=0$ if $j>0$, $x(dx)^{l-1}x=0$ for $l>1$ and 
$\int_\RR ds\, \eva_0 (p(\delta p)^2) = 
2\pi\imath \;\langle\#_\tau\Tr^s,p\rangle = -\,\imath \;
\Tr(\rho(p))$,
we get
\begin{eqnarray*}
\!\!\!\!\!\!
& & 
\pair{\#_{\tau\otimes \idl}\,\eta^s}{\Phi^{-1}(X)}\;=\; 
c_{2k+2}\int_{\tau \otimes \idl}^s
\Phi^{-1}(X (d^s_{\tau\otimes\idl}X)^{2k+2})\\
\!\!\!\!\!\!
&  &\;\;\; =\; 
c_{2k+2}\sum_{0<l<j\leq 2k+2}\int_\RR ds\, \eva_0 \left(p
(\delta p) p^{j-l-1} (\delta p)\right) 
\int x(d x)^{l-1}x(d x)^{j-i-1}x(d x)^{2k+2-j}\\
\!\!\!\!\!\!
&  &\;\;\; =\; 
 -\,\imath\;\frac{(k+1)c_{2k+2}}{c_{2k}} \;\Tr(\rho(p)) 
\;\langle\eta,x \rangle \\
\!\!\!\!\!\!
&  &\;\;\; =\; 
 -\,\frac{1}{2\pi}\; \langle\Tr\otimes\eta,X \rangle\;.
\end{eqnarray*}


Next we apply this to a unitary 
$X=\rho(p)\otimes x + \rho(p)^\perp\otimes 1$
for $x\in U(\Bb)$ and a projection $p$.
Then
$d_{ \tau\otimes \idl}^s(\Phi^{-1}(X)) = p\otimes d x +
\delta  p \otimes (x-1)$.
If $n=2k+1$, $k>0$, one obtains, 
taking into account that  $p(\delta p)p^k=0$ for $k>0$,
\begin{eqnarray*}
\!\!\!\!\!\!\!\!\!
&  &
\pair{\#_{\tau\otimes \idl}\,\eta^s}{\Phi^{-1}(X)}
\;=\; 
c_{2k+3}\int_{\tau \otimes \idl}^s
\Phi^{-1}((X^*-1)d^s_{\tau\otimes\idl}X
(d^s_{\tau\otimes\idl}X^* d^s_{\tau\otimes\idl}X)^{k+1})\\
\!\!\!\!\!\!\!\!\!
&  & =
c_{2k+3}\int_\RR ds\, \eva_0 \left(p (\delta p)^2\right)
\sum_{0<j< 2k+3}
\int (x^*-1)d x^{(1)}\cdots dx^{(j-1)}(x^*-1)(x-1)dx^{(j+2)}\cdots d x^{(2k+3)}
,
\end{eqnarray*}
where $x^{(j)}=x$ if $j$ is odd and $x^{(j)}=x^*$ if $j$ is even.
We claim that

\begin{eqnarray} \label{eq1}
&& \sum_{j=1}^{2k+2}
\int (x^*-1)d x^{(1)}\cdots dx^{(j-1)}(x^*-1)(x-1)dx^{(j+2)}
\cdots d x^{(2k+3)}\\
&& \qquad\qquad  = 2(2k+3)\int (x^*-1)d x (d x^* dx)^{k}, 
\nonumber
\end{eqnarray} which then implies
$$
\pair{\#_{\tau\otimes \idl}\,\eta^s}{\Phi^{-1}(X)}
\; =\; 
 -\,\imath\; \frac{2(2k+3)c_{2k+3}}{c_{2k+1}}\;
\Tr(\rho(p)) \langle\eta,x \rangle
\; =\; 
 -\,\frac{1}{2\pi}\;\langle\Tr\otimes\eta,X \rangle\;.
$$
Since $(x^*-1)(x-1)=2 - x - x^*$, equation (\ref{eq1}) is equivalent to
\begin{equation}\label{eq2} \int\sum_{l=0}^k (x^*-1)(dx dx^*)^l
\left((x+x^*)dx + dx (x+x^*)\right)
(dx^* dx)^{k-l} =  
-2\int (x^*-1)(d x dx^*)^{k}dx\;. 
\end{equation}
Now use $dx^* x = - x^* dx$ and $dx x^* = - x dx^*$ to pull $x$
and $x^*$ of $(x+x^*)$ either to the right or to the left and then use
cyclicity in order to obtain 
\begin{eqnarray*}
\mbox{l.h.s. of (\ref{eq2})} &=& \sum_{l=0}^{k} 
\int x(x^*-1)\left( (dx^* dx)^l dx (dx^* dx )^{k-l} 
-(dx dx^*)^l dx^* (dxdx^*)^{k-l}\right. \\ 
& & \qquad \mbox{ } \qquad\mbox{ } \qquad\left.  
-(dx^* dx)^l dx^* (dx^*dx)^{k-l} + (dx dx^*)^l dx (dx dx^* )^{k-l}
\right) \\
&=& \sum_{l=0}^{k}
\left(-\,\eta(x,\underbrace{x^*,x,\dots}_{2l},x,
\underbrace{x^*,x,\dots}_{2(k-l)})
\,+\,\eta(x,\underbrace{x,x^*,\dots}_{2l},x^*,
\underbrace{x,x^*,\dots}_{2(k-l)})\right.\\
& & \qquad \mbox{ } \quad \left.
+\,\eta(x,\underbrace{x^*,x,\dots}_{2l},x^*,\underbrace{x^*,x,\dots}_{2(k-l)})
\,-\,\eta(x,\underbrace{x,x^*,\dots}_{2l},
x,\underbrace{x,x^*,\dots}_{2(k-l)})\right)\;
.
\end{eqnarray*}
Here the entries under-braced are alternating. For fixed $l$ the first
and the fourth term in each summand cancel by cyclic symmetry.
The remaining terms are
\begin{eqnarray*} \mbox{l.h.s. of (\ref{eq2})}
& = & \sum_{l=0}^{k}
\left(-\,\eta(\underbrace{x,x^*,\dots}_{2l},
\underbrace{x^*,x,\dots}_{2(k-l+1)})
\,+\,\eta(\underbrace{x,x^*,\dots}_{2(l+1)},
\underbrace{x^*,x,\dots}_{2(k-l)})
\right)\\
& = & -\,\eta(\underbrace{x^*,x\dots}_{2(k+1)})
\,+\,\eta(\underbrace{x,x^*,\dots}_{2(k+1)}) \\
& = & -\,\int 2(x^*-1) dx (dx^* dx)^{k}\;.
\end{eqnarray*}

It remains to compute the pairings for unitaries $X$ of the form
$X-1=\sum_j a_j\otimes b_j\in \K\otimes\Bb$ where the sum is finite and
the $a_j$ have finite rank (for projections of the form 
$\sum_j a_j\otimes b_j\in \K\otimes\Bb^+$ the argument is similar).
In that case $X-1\in M_n\otimes\Bb$
($M_n=M_n(\CC)$ is the associated sub algebra of $\K$ for some
finite $n$). Let $e\in M_n$ be an arbitrary rank one projection, $e^\perp$ its
ortho complement in $M_n$. Then, by the above,
$$
\langle\Tr\otimes\eta,X\rangle 
\;= \;\langle
\Tr\otimes\Tr\otimes\eta,e\otimes X + e^\perp\otimes 1\rangle
\;=\;  -\,2\pi\;\langle \#_{\tau\otimes\idl}(\Tr\otimes\eta)^s,
\rho^{-1}(e)\otimes X+\rho^{-1}(e^\perp)\otimes 1\rangle\,.$$
Further let
$U\in \Uu(M_n\otimes M_n)$ be a unitary such that 
$\mbox{\rm Ad}_U\in\mbox{\rm Aut}(M_n\otimes
M_n)$ is the flip, $U a_1\otimes a_2 U^*= a_2\otimes a_1$. Since 
$\Uu(M_n\otimes M_n)$ is connected, a path connecting $U$ to the
identity gives rise to a homotopy in $M_n\otimes M_n\otimes\Bb^+$
between  $e\otimes X+ e^\perp\otimes 1$ and 
$\mbox{\rm Ad}_U\otimes\id(e\otimes X+ e^\perp\otimes 1)$.
Since $\mbox{\rm Ad}_U\otimes\id(e\otimes X)=\sum_j
a_j\otimes e\otimes b_j $, we have by homotopy invariance of the
pairings 
\begin{eqnarray*} 
&&\langle \#_{\tau\otimes\idl}{(\Tr\otimes\eta)}^s,
\rho^{-1}(e)\otimes X+\rho^{-1}(e^\perp)\otimes 1 \rangle\\
&& \qquad \qquad  \qquad \qquad = 
\langle \#_{\tau\otimes\idl}(\Tr\otimes\eta)^s,\sum_j a_j\otimes
\rho^{-1}(e)\otimes b_j +  
\mbox{\rm Ad}_U\otimes\id(e^\perp\otimes 1) \rangle \\
&& \qquad \qquad  \qquad \qquad = 
\langle \#_{\tau\otimes\idl}{\eta}^s,\sum_j a_j\otimes b_j+1\rangle
\;.
\end{eqnarray*}
This proves the statement.
\eb

\vspace{.2cm}

\noindent 
{\bf Proof of Theorem~\ref{gleich2}} Combine the last two lemmas with 
$X=\Phi\Psi(x)$. 
\eb

\end{appendix}

\end{document}